\documentclass[prd,draft,amsfonts,amssymb,amsmath,showpacs]{revtex4}

\begin{document}

\title{Algebra of Noncommutative Riemann Surfaces}
\author{Tadafumi Ohsaku}

\date{\today}


\begin{abstract}

We examine several algebraic properties of the noncommutive $z$-plane and Riemann surfaces. 
The starting point of our investigation is a two-dimensional noncommutative field theory,
and the framework of the theory will be converted into that of a complex coordinate system.
The basis of noncommutative complex analysis is obtained thoroughly, and the 
considerations on functional analysis are also given before performing the examination 
of the conformal mapping and the Teichm\"{u}ller theory.
( Keywords; Complex Analysis, Riemann Surfaces and Teichm\"{u}ller Space, Functional Analysis, 
Deformation Quantization, Non-Commutative Geometry, Quantum Groups )

\end{abstract}

\pacs{02.20.Tw, 02.20.Uw, 02.40.Gh, 02.40.Tt, 11.25.-w}

\maketitle

{\it Wovon man nicht sprechen kann, dar\"{u}ber muss man schweigen.}
( Ludwig Wittgenstein, {\it Tractatus Logico-Philosophicus}. )

\section{Introduction}

Recently, various works have been done on theories of noncommutative
geometry, by mathematician~[1-11] and mathematical/theoretical physicists~[12-14].
Connes emphasized that, several standard methods of calculus and analysis could not
be employed to noncommutative cases~[1].
Hence it is important problem for us to construct a noncommutative model
and reveal it in detail, closed with several contexts of mathematical/theoretical physics.
In this paper, we will examine the basis of noncommutative complex analysis.
The noncommutativity realizes on the complex $z$-plane and Riemann surfaces~[15-22].
We will show that, the field of complex numbers ${\bf C}$ in the ordinary case
obtains various exotic properties after the introduction of the noncommutativity. 
In fact, complex numbers do not
satisfy the criteria of the field ${\bf C}$ in the exact sense.
The definition for the concept of Riemann surfaces will be chosen as follows~[17-19]:
$\Sigma$ is called as a Riemann surface, when a system of local complex coordinates
gives mappings they are homeomorphic and biholomorphic, and it is defined on a 
connected Hausdorff space $\Sigma$.
The noncommutative complex plane is, in fact, a Hausdorff topological space,
and noncommutative Riemann surfaces permit Poisson, symplectic and complex structures.
We use the word "mapping" as the sense for differentiable manifolds.
A function $f(z,\bar{z})$ obeys the noncommutative algebra coming from the noncommutativity
of complex coordinates $z$ and $\bar{z}$. 
A mapping ( or, a morphism ) $\varphi$ we consider here has its inverse $\varphi^{-1}$.
On the other hand, as we will show later, a holomorphic function $f(z)$ always does not have its inverse 
in the algebraic sense $f(z)f(z)^{-1}\ne f(z)^{-1}f(z)=1$.
Several papers of Riemann surfaces with noncommutativity in the context of quantum group
of Drinfel'd and Woronowicz, to construct examples of quantum space, 
were given in literature~[23-25], similar to the noncommutative sphere. 
On the contrary, this paper can be entitled as "Algebra of Riemann surfaces of two-dimensional
noncommutative field theory".


This paper is organized as follows.
In Sec. II, we will give the two-dimensional noncommutative theory in ${\bf R}^{2}$,
as the starting point of our study. 
In Sec. III, we will convert it into the case on complex coordinate systems, 
and examine how the methods and various results of complex analysis are changed 
in the noncommutative $z$-plane and Riemann surfaces.
For obtaining knowledge on topological structures of the noncommutative theory 
from another point of view, we discuss the methods of functional analysis in Sec. IV.
Conformal mappings and algebra of the noncommutative $z$-plane and 
Riemann surfaces will be examined in Sec. V.
Finally in Sec. VI, summary with further possible extensions of 
this work will be discussed.

\section{The two-dimensional Noncommutativity}

In this section, we follow the discussions of ${\it so}$-${\it called}$ noncommutative field theory~[12-14],
and we introduce a two-dimensional cartesian coordinate system $\sigma_{a}\in{\bf R}^{2}$ ( $a=1,2$ )
with requiring the following noncommutativity:
\begin{eqnarray}
& & [\sigma_{a},\sigma_{b}] = iH_{ab} = -iH_{ba} =
\left(
\begin{array}{cc}
0 & iH \\
-iH & 0
\end{array}
\right)_{ab}, \quad \sigma_{1}\perp\sigma_{2}. 
\end{eqnarray}
The noncommutativity parameter $H$ is a $c$-number ( complex in general ), 
global and independent on any coordinates of ${\bf R}^{2}$. 
Therefore one finds
\begin{eqnarray}
& & \sigma'_{a} = \sigma_{a}+ d\sigma_{a},  \quad
\sigma'_{b} = \sigma_{b}+ d\sigma_{b},  \nonumber \\
& & [\sigma'_{a},\sigma'_{b}] = [\sigma_{a},\sigma_{b}] \to 
[ d\sigma_{a},\sigma_{b} ] + [ \sigma_{a}, d\sigma_{b} ] 
+ [ d\sigma_{a}, d\sigma_{b} ] = 0.
\end{eqnarray}
Furthermore, the relations $[ d\sigma_{a},\sigma_{b} ]=0$ and $[ d\sigma_{a}, d\sigma_{b} ]=0$
will be assumed for constructing the theory of noncommutative complex analysis.
Because of the noncommutativity (1), a product of functions of $\sigma_{a}$ has to be ordered.
We choose the Weyl-ordering. For example,
\begin{eqnarray}
(\sigma_{1}\sigma_{2})_{W} &=& \frac{1}{2}(\sigma_{1}\sigma_{2}+\sigma_{2}\sigma_{1}), \quad
(\sigma_{1}\sigma^{2}_{2})_{W} = \frac{1}{3}
(\sigma_{1}\sigma^{2}_{2}+\sigma_{2}\sigma_{1}\sigma_{2}+\sigma^{2}_{2}\sigma_{1}).
\end{eqnarray}
Because $\sigma_{a}$ are Hermitian, their Weyl-ordered products are always Hermitian.
The definition of the star product of functions of $\sigma_{a}$ is given as follows:
\begin{eqnarray} 
f(\sigma) &\equiv& f(\sigma_{1},\sigma_{2}),  \quad g(\sigma) \equiv g(\sigma_{1},\sigma_{2}),  \nonumber \\
F(\sigma) &\equiv& f(\sigma)\star g(\sigma)    \nonumber \\
&=& f(\sigma)\exp\Bigl(
i\frac{H_{ab}}{2}
\frac{\overleftarrow{\partial}}{\partial\sigma_{a}}\frac{\overrightarrow{\partial}}{\partial\sigma_{b}}
\Bigr)g(\sigma) = f(\sigma)g(\sigma) + \frac{i}{2}H\{f,g\}^{P.B.}_{\sigma_{1},\sigma_{2}} + \Xi[f,g]  \nonumber \\
&=& \sum^{\infty}_{n=0}F_{(n)}(\sigma)H^{n},  
\end{eqnarray}
where,
\begin{eqnarray}
\{f,g\}^{P.B.}_{\sigma_{1},\sigma_{2}} &=& 
\frac{\partial f(\sigma)}{\partial \sigma_{1}}\frac{\partial g(\sigma)}{\partial \sigma_{2}} -
\frac{\partial f(\sigma)}{\partial \sigma_{2}}\frac{\partial g(\sigma)}{\partial \sigma_{1}},   \\
\Xi[f,g] &=& \sum^{\infty}_{n=2}\frac{1}{n!}\Bigl(\frac{iH}{2}\Bigr)^{n}f(\sigma)
\Bigl( \frac{\overleftarrow{\partial}}{\partial\sigma_{1}}\frac{\overrightarrow{\partial}}{\partial\sigma_{2}}
-\frac{\overleftarrow{\partial}}{\partial\sigma_{2}}\frac{\overrightarrow{\partial}}{\partial\sigma_{1}}\Bigr)^{n}g(\sigma).
\end{eqnarray}
The star product $F(\sigma)$ is formally defined as a power series of $H$, 
and in fact it is an entire function of $H$ inside the convergence radius of the series.
The convergence radius of the series depends on $f(\sigma)$ and $g(\sigma)$,
and we have to determine them for the estimation of the radius.
$\{f,g\}^{P.B.}_{\sigma_{1},\sigma_{2}}$ is a Poisson bracket, 
and it introduces the quasi Poisson structure to a theory.
The star product will be rewritten in the following form:
\begin{eqnarray}
f(\sigma)\star g(\sigma) &=& f(\sigma)\exp
\Bigl[ 
\frac{i}{2}H\overleftarrow{\partial}_{\Sigma}{\cal I}\overrightarrow{\partial}_{\Sigma}  
\Bigr]g(\sigma),    \nonumber \\
\overleftarrow{\partial}_{\Sigma} &\equiv& 
\Bigl(\frac{\overleftarrow{\partial}}{\partial\sigma_{1}},\frac{\overleftarrow{\partial}}{\partial\sigma_{2}}\Bigr),  \quad
\overrightarrow{\partial}_{\Sigma} \equiv 
\Bigl(\frac{\overrightarrow{\partial}}{\partial\sigma_{1}},\frac{\overrightarrow{\partial}}{\partial\sigma_{2}}\Bigr)^{T},  \quad
{\cal I} \equiv \left(
\begin{array}{cc}
0 & 1 \\
-1 &  0
\end{array}
\right).
\end{eqnarray}
Hence, the star product shows the symplectic structure. 
The star products do not commute, and the associativity is broken ${\it in}$ ${\it general}$:
\begin{eqnarray}
f(\sigma)\star g(\sigma) &\ne& g(\sigma)\star f(\sigma),  \\
f(\sigma)\star \bigl( g(\sigma)\star h(\sigma) \bigr) &\ne& \bigl( f(\sigma)\star g(\sigma) \bigr)\star h(\sigma),
\end{eqnarray}
while, the star product satisfies the linear algebraic relation:
$f\star(xg+yh) = x(f\star g) + y(f\star h)$ $(x,y\in{\bf C})$.
Several equations for the Poisson brackets are summarized as follows:
\begin{eqnarray}
\{f+g,h\}^{P.B.}_{\sigma_{1},\sigma_{2}} &=& 
\{f,h\}^{P.B.}_{\sigma_{1},\sigma_{2}}+\{g,h\}^{P.B.}_{\sigma_{1},\sigma_{2}},  \quad
\{xf,yg\}^{P.B.}_{\sigma_{1},\sigma_{2}} = xy\{f,g\}^{P.B.}_{\sigma_{1},\sigma_{2}},  \nonumber  \\
\{f,f\}^{P.B.}_{\sigma_{1},\sigma_{2}} \pm \{g,g\}^{P.B.}_{\sigma_{1},\sigma_{2}} &=& 
\frac{1}{2}\Bigl( \{f+g,f\pm g\}^{P.B.}_{\sigma_{1},\sigma_{2}} 
+ \{f-g,f\mp g\}^{P.B.}_{\sigma_{1},\sigma_{2}}  \Bigr),   \nonumber  \\
\{g,f\}^{P.B.}_{\sigma_{1},\sigma_{2}} \pm \{f,g\}^{P.B.}_{\sigma_{1},\sigma_{2}} &=& 
\frac{1}{2}\Bigl( \{f+g,f\pm g\}^{P.B.}_{\sigma_{1},\sigma_{2}} 
- \{f-g,f\mp g\}^{P.B.}_{\sigma_{1},\sigma_{2}}  \Bigr) 
= \Bigl[ \frac{\partial g}{\partial\sigma_{1}}, \frac{\partial f}{\partial\sigma_{2}}  \Bigr]_{\mp}
- \Bigl[ \frac{\partial g}{\partial\sigma_{2}}, \frac{\partial f}{\partial\sigma_{1}}  \Bigr]_{\mp}.
\end{eqnarray}
Here, $[\quad,\quad]_{-}$ ( $[\quad,\quad]_{+}$ ) denotes a (anti)commutator.
The axiom of Poisson manifolds, namely both the Jacobi identity
$0= \{ \{f,g\}^{P.B.}_{\sigma_{1},\sigma_{2}}, h \}^{P.B.}_{\sigma_{1},\sigma_{2}}
+\{ \{g,h\}^{P.B.}_{\sigma_{1},\sigma_{2}}, f \}^{P.B.}_{\sigma_{1},\sigma_{2}}
+\{ \{h,f\}^{P.B.}_{\sigma_{1},\sigma_{2}}, g \}^{P.B.}_{\sigma_{1},\sigma_{2}}$,
and the skew symmetry $\{f,g\}^{P.B.}_{\sigma_{1},\sigma_{2}}=-\{g,f\}^{P.B.}_{\sigma_{1},\sigma_{2}}$ 
are not satisfied ${\it in}$ ${\it general}$.

\section{Complex Analysis}

For the investigation of the noncommutativity in complex manifolds,
we introduce a complex coordinate system of a flat space ( the Gaussian plane ),
with referring the results of the previous section:
\begin{eqnarray}
z &=& \sigma_{1}+i\sigma_{2}, \quad  \bar{z} = \sigma_{1}-i\sigma_{2}, \nonumber \\
dz &=& d\sigma_{1}+id\sigma_{2}, \quad d\bar{z} = d\sigma_{1}-id\sigma_{2}, \quad 
ds^{2} = dzd\bar{z} = d\bar{z}dz = d^{2}\sigma_{1}+d^{2}\sigma_{2},   \nonumber \\
\partial_{z} &=& \frac{1}{2}\Bigl(\frac{\partial}{\partial\sigma_{1}}-i\frac{\partial}{\partial\sigma_{2}}\Bigr), \quad 
\partial_{\bar{z}} = \frac{1}{2}\Bigl(\frac{\partial}{\partial\sigma_{1}}+i\frac{\partial}{\partial\sigma_{2}}\Bigr).
\end{eqnarray}
In this case, all of the Christoffel symbols vanish.
In the ordinary commutative case of a complex manifold, 
we can choose the conformally flat isothermal metric as follows:
\begin{eqnarray}
g_{z\bar{z}} &=& g_{\bar{z}z} = \frac{1}{2}e^{2\eta(z,\bar{z})}, \quad g_{zz} = g_{\bar{z}\bar{z}} = 0,  \nonumber \\
g^{z\bar{z}} &=& g^{\bar{z}z} = 2e^{-2\eta(z,\bar{z})}, \quad g^{zz} = g^{\bar{z}\bar{z}} = 0,  \nonumber \\
ds^{2} &=& g_{z\bar{z}}dzd\bar{z} + g_{\bar{z}z}d\bar{z}dz = e^{2\eta(z,\bar{z})}dzd\bar{z}.
\end{eqnarray}
However, if we choose this metric, we have to consider how to handle the Weyl rescaling factor 
$e^{2\eta(z,\bar{z})}$, and it is a nontrivial problem for us to consider 
an Hermite-K\"{a}hler manifold of the noncommutative $z$-plane.
Later, we examine the noncommutativity of generic situations of Riemann surfaces
through the uniformization theorem or the Riemann's mapping theorem 
( the Gaussian plane ${\bf C}$, the unit disc ${\cal D}\equiv\{z\in{\bf C}|1>|z|\}$, 
the Riemann sphere $\widehat{\bf C}={\bf C}\bigcup\{\infty\}$, 
a torus ${\bf C}/\Gamma$ with a lattice group $\Gamma$, 
the upper half ${\bf H}$ and the lower half $\overline{\bf H}$ of the $z$-plane , ... ),
though a large part of the results of this paper have nothing to do with a choice of domain $\Omega$.
It is possible for us to consider the case of genus $g\ge 2$ Riemann surface under the noncommutativity. 
Constructions of complex projective spaces and algebraic varieties 
from our starting point has difficulties coming from the noncommutativity of $z$ and $\bar{z}$.
We can consider a bijective mapping $\varphi: M_{c}\to M_{nc}$
for introducing the noncommutativity into our theory, 
where $M_{c}$ corresponds to the case of the usual commutative $z$-plane ( or, a Riemann surface ), 
while $M_{nc}$ denotes the noncommutative $z$-plane ( or, a noncommutative Riemann surface ).
This correspondence $\varphi$ conserves a part of several characters of $M_{c}$
while various properties will change in $M_{nc}$.
The purpose of this paper is dedicated for the examination of them.
The Laplacian becomes $\Delta = \partial^{2}/\partial\sigma^{2}_{1}+\partial^{2}/\partial\sigma^{2}_{2}
=4\partial_{z}\partial_{\bar{z}}$.


Now, the noncommutativity (1) is translated into
\begin{eqnarray}
& & [z,\bar{z}] = 2H, \quad [z,z] = [\bar{z},\bar{z}] = 0.
\end{eqnarray}
Note that the commutator $[z,\bar{z}]$ is given by the holomorphic function $z$ and 
the antiholomorphic function $\bar{z}$.
Thus, the noncommutativity nontrivially relates arbitrary two points they symmetrically locate 
the upper ( ${\bf H};\Im(z)>0$ ) and lower ( $\overline{\bf H};\Im(z)<0$ ) halves 
of the Gaussian $z$-plane ( the complex conjugation ), and the origin of the $z$-plane is fixed by (13): 
The translation invariance is broken by introducing the commutator.
We regard that the commutator $[z,\bar{z}]=2H$ is satisfied at the origin $z=\bar{z}=0$ of ${\bf C}$.
The star product of functions given by the complex coordinates becomes
\begin{eqnarray}
f(\sigma)\star g(\sigma) &=& 
f(\sigma)\exp\Bigl(
i\frac{H_{ab}}{2}
\frac{\overleftarrow{\partial}}{\partial\sigma_{a}}\frac{\overrightarrow{\partial}}{\partial\sigma_{b}}
\Bigr)g(\sigma)   \nonumber  \\
&=& f(z,\bar{z})\exp\Bigl[H (
\overleftarrow{\partial}_{z}\overrightarrow{\partial}_{\bar{z}}-
\overleftarrow{\partial}_{\bar{z}}\overrightarrow{\partial}_{z}
) \Bigr]g(z,\bar{z}) = f(z,\bar{z})g(z,\bar{z}) + H\{f,g\}^{P.B.}_{z,\bar{z}} + \cdots,
\end{eqnarray}
where,
\begin{eqnarray}
\{f,g\}^{P.B.}_{z,\bar{z}} &\equiv& 
\frac{\partial f}{\partial z}\frac{\partial g}{\partial \bar{z}}-
\frac{\partial f}{\partial \bar{z}}\frac{\partial g}{\partial z},  \nonumber \\
\{z,z\}^{P.B.}_{z,\bar{z}} &=& \{\bar{z},\bar{z}\}^{P.B.}_{z,\bar{z}} = 0, \quad
\{z,\bar{z}\}^{P.B.}_{z,\bar{z}} = -\{\bar{z},z\}^{P.B.}_{z,\bar{z}} = 1,   \nonumber \\
\{z^{l},z^{m}\}^{P.B.}_{z,\bar{z}} &=& \{\bar{z}^{l},\bar{z}^{m}\}^{P.B.}_{z,\bar{z}} = 0, \quad
\{z^{l},\bar{z}^{m}\}^{P.B.}_{z,\bar{z}} = -\{\bar{z}^{m},z^{l}\}^{P.B.}_{z,\bar{z}} = lmz^{l-1}\bar{z}^{m-1}, \quad ( \forall l,m\in {\bf N} ). 
\end{eqnarray}
The star product is rewritten in the following form
where a symplectic structure is clearly shown:
\begin{eqnarray}
f(z,\bar{z})\star g(z,\bar{z}) &=&
f(z,\bar{z})\exp\Bigl[ H\overleftarrow{\partial}_{Z}{\cal I}\overrightarrow{\partial}_{Z} \Bigr]g(z,\bar{z}),   
\quad
\overleftarrow{\partial}_{Z} \equiv (\overleftarrow{\partial}_{z},\overleftarrow{\partial}_{\bar{z}}),  \quad
\overrightarrow{\partial}_{Z} \equiv (\overrightarrow{\partial}_{z},\overrightarrow{\partial}_{\bar{z}})^{T}.
\end{eqnarray}
We have observed the Poisson structure of the noncommutative complex coordinate system. 
The set of $z$ and $\bar{z}$ is regarded as a kind of canonical coordinates,
and we can determine a canonical transformation through the Poisson brackets. 
Therefore, 
a $U(1)\sim O(2)$ group keeps the ( local or global ) Euclid structure coming from the two-dimensional
coordinate system $(\sigma_{1},\sigma_{2})$, 
operations of $GL(1,{\bf C})$ keep the complex structure $z=\sigma_{1}+i\sigma_{2}$,
and a canonical transformation $Sp(1,{\bf C})$ keeps the symplectic structure
coming from the definition of the star product. 
A canonical transformation $(z,\bar{z})\to(z',\bar{z}')$ satisfies
\begin{eqnarray}
\{z,z\}^{P.B.}_{z,\bar{z}} &=& \{z',z'\}^{P.B.}_{z,\bar{z}} 
= \{\bar{z},\bar{z}\}^{P.B.}_{z,\bar{z}} = \{\bar{z}',\bar{z}'\}^{P.B.}_{z,\bar{z}} = 0, \nonumber \\
\{z,\bar{z}\}^{P.B.}_{z,\bar{z}} &=& \{z',\bar{z}'\}^{P.B.}_{z,\bar{z}} 
= -\{\bar{z},z\}^{P.B.}_{z,\bar{z}} = -\{\bar{z}',z'\}^{P.B.}_{z,\bar{z}} = 1. 
\end{eqnarray}
The following unimodular transformation gives an example of the canonical transformations:
\begin{eqnarray}
\left( 
\begin{array}{c}
z'  \\
\bar{z}'
\end{array}
\right) &=& \left(
\begin{array}{cc}
a & b \\
c & d
\end{array}
\right)\left(
\begin{array}{c}
z \\
\bar{z}
\end{array}
\right), \quad ad-bc =1.
\end{eqnarray}


We can use several concepts of classical mechanics~[26].
If we introduce the following 1-form $\omega_{1}$ and 2-form $\omega_{2}$,
\begin{eqnarray}
\omega_{2} &=& d\sigma_{1}\wedge d\sigma_{2} 
= \frac{\partial(\sigma_{1},\sigma_{2})}{\partial(z,\bar{z})}dz\wedge d\bar{z}
= \frac{i}{2}dz\wedge d\bar{z}, \nonumber \\
\omega_{1} &=& \frac{i}{2}zd\bar{z}, \quad \overline{\omega_{1}} = -\frac{i}{2}\bar{z}dz, \quad
\overline{\omega_{2}} = -\frac{i}{2}d\bar{z}\wedge dz = \omega_{2},  \quad 
d = dz\wedge\partial_{z}+d\bar{z}\wedge\partial_{\bar{z}}, \quad d\omega_{2} = 0,
\end{eqnarray}  
and a canonical transformation $(z,\bar{z})\to(z',\bar{z}')$ not only satisfies 
$dz\wedge d\bar{z}=dz'\wedge d\bar{z}'$, 
but also keeps a volume of the "phase space" ( the Liouville theorem ):
\begin{eqnarray}
\int_{D} dz\wedge d\bar{z} &=& \int_{D'} dz'\wedge d\bar{z}'.
\end{eqnarray}
Later, we examine integrations of functions of the noncommutative $z$-plane.
From the definition of the star product, one finds
\begin{eqnarray}
{\cal F}(z,\bar{z},H) &\equiv& f(z,\bar{z})\star g(z,\bar{z}),   \quad
\frac{d}{dH}{\cal F}(z,\bar{z},H)\Big|_{H=0} = \{f,g\}^{P.B.}_{z,\bar{z}}.
\end{eqnarray}
${\cal F}(z,\bar{z},H)$ might be regarded as a function of the two-dimensional ( complex )
phase space ${\cal P}(\sigma_{1},\sigma_{2})={\cal P}_{c}(z,\bar{z})$, supplemented by the parameter $H$. 
The above differential equation has been given as a Lie derivative.
Hence, one can introduce a Hamilton vector field in our theory~[26].
If we define a Hamilton vector field $X_{f}$ associated with a function $f(z,\bar{z})$ 
in the following form, the Poisson bracket is expressed as
\begin{eqnarray}
X_{f} &\equiv& 
\frac{\partial f}{\partial z}\frac{\partial}{\partial\bar{z}}
-\frac{\partial f}{\partial\bar{z}}\frac{\partial}{\partial z},  \quad
\{ f,g\}^{P.B.}_{z,\bar{z}} = X_{f}g,
\end{eqnarray}
though, the Poisson bracket could not be expressed as a skew scalar product $\langle X_{f},X_{g}\rangle$ of
a symplectic manifold because of the noncommutativity of $f(z,\bar{z})$ and $g(z,\bar{z})$ in general.
By the definition, we obtain the following canonical form of "equations of motion": 
\begin{eqnarray}
X_{f}z &=& \{f,z \}^{P.B.}_{z,\bar{z}} = -\frac{\partial f}{\partial \bar{z}}, \quad
X_{f}\bar{z} = \{f,\bar{z}\}^{P.B.}_{z,\bar{z}} = \frac{\partial f}{\partial z}.
\end{eqnarray}


We use the following definition for norm of the function $z$:
\begin{eqnarray}
z &=& |z|\exp\bigl\{i (\arg z)\bigr\}, \quad 
|z| \equiv \sqrt{\sigma^{1}_{1}+\sigma^{2}_{2}}.  
\end{eqnarray}
$|z|$ gives the distance between a point $z$ and the origin of the $z$-plane.
In this definition, $|\bar{z}|=|z|$ is satisfied.
For the unit disc ${\cal D}$, we have choices of metric as follows:
The Euclidean metric $ds^{2}=dzd\bar{z}$, or the Poincar\'{e} metric,
\begin{eqnarray}
ds^{2} &=& \frac{4}{(1-|z|^{2})^{2}}dzd\bar{z}  
= \frac{4}{1-2(\sigma^{2}_{1}+\sigma^{2}_{2})
+ \sigma^{4}_{1} + \sigma^{2}_{2} + \sigma^{2}_{1}\sigma^{2}_{2} +\sigma^{2}_{2}\sigma^{2}_{1}}
(d^{2}\sigma_{1}+d^{2}\sigma_{2}).
\end{eqnarray} 
Here, the Poincar\'{e} metric of the noncommutative case has ambiguities coming from $[\sigma_{1},\sigma_{2}]=iH$.
In fact, there are ambiguities of definitions of distances of noncommutative Riemann surfaces
by using several Riemannian metrics.
( The Poincar\'{e} metric for ${\bf H}$, $ds^{2}=1/(\Im z)^{2}dzd\bar{z}=1/\sigma^{2}_{2}dzd\bar{z}$
has no ambiguity. )
However, we can define a distance of the noncommutative $z$-plane by
\begin{eqnarray}
\rho(z(t_{1}),z(t_{2})) &\equiv& \inf\Bigl[ \int^{t_{2}}_{t_{1}}|z'(t)|dt\Bigg| |z'(t)|=\frac{d|z(t)|}{dt}  \Bigr],
\end{eqnarray}
where $t$ is a real parameter $t_{1}\le t \le t_{2}$, 
$z(t)$ is assumed as a $C^{1}$ curve under taking derivative with respect to $t$, 
and $|z(t)|\equiv\sqrt{\sigma^{2}_{1}(t)+\sigma^{2}_{2}(t)}$.
Thus, if the distance behaves as a Cauchy sequence $\rho(z(t_{m}),z(t_{n}))\to 0$ under $n,m\to\infty$,
$(M_{nc},\rho)$ is complete.
From the definition, one finds
\begin{eqnarray}
\sigma_{1} = \frac{|z|}{2}(e^{i\arg z} + e^{-i\arg z}), \quad
\sigma_{2} = \frac{|z|}{2i}(e^{i\arg z} - e^{-i\arg z}).
\end{eqnarray}
Hence, the commutator will be expressed by an argument of $z$:
\begin{eqnarray}
2H = [z,\bar{z}] = |z|^{2}[e^{i\arg z}, e^{-i\arg z}], \quad \lim_{|z|\to +\infty}[e^{i\arg z}, e^{-i\arg z}] = \lim_{|z|\to +\infty}\frac{2H}{|z|^{2}} = 0.
\end{eqnarray}
Here, the noncommutativity is carried by the argument of $z$.
At $|z|\to \infty$, $z$ and $\bar{z}$ commute with each other:
The effect of the noncommutativity vanishes at $|z|\to\infty$ ( or, at $z=\infty$ in $\widehat{\bf C}$ ).
On the other hand, by the mapping $\varphi: z \longleftrightarrow 1/z (\equiv\bar{z}/|z|^{2})$,
\begin{eqnarray}
&& \Bigl[z,\frac{1}{z}\Bigr] = 2H|z|^{-2} \stackrel{\varphi}{\longrightarrow} \Bigl[\frac{1}{z},z\Bigr] = 2H|z|^{2},
\end{eqnarray}
thus,
\begin{eqnarray}
&& [z,z^{-1}] = 2H|z|^{-2} = -2H|z|^{2}.
\end{eqnarray}
Therefore, we have found that the sign of the noncommutativity parameter $H$ relates 
to the complex conjugation $\rho: z\leftrightarrow \bar{z}$.
The equation $z^{-1}=1/z=\bar{z}/|z|^{2}$ has to be regarded as the definition of $z^{-1}$,
because we would like to ( must ) remove the ambiguity for the definition coming from 
$\bar{z}/(z\bar{z})\ne\bar{z}/(\bar{z}z)$.
In other words, the commutator $[z,z^{-1}]=2H|z|^{-2}$ determines the algebra which $z^{-1}$ obeys.
$|z|$ includes both $\sigma_{1}$ and $\sigma_{2}$, and then the algebraic relation of $|z|$ ( $|z|^{2}$ ) 
and $f(z)$ is unclear if we consider it naively.
We cannot take $z(z^{-1})=1$ or $(z^{-1})z=1$, because $z$ and $z^{-1}$ are not trivial $c$-numbers
under the relation with each other ( $z\bar{z}\ne\bar{z}z\ne|z|^{2}$ ),
though $z$, $\bar{z}$, $z^{-1}$ and $\bar{z}^{-1}$ cannot be regarded as simple operators.
Under our definition, $|1/z|=|z|^{-1}=|\bar{z}|/|z|^{2}$ and $|z||z|^{-1}=|z|^{-1}|z|=1$.
Therefore, one gets
\begin{eqnarray}
&& [z,\bar{z}^{n}] = 2nH\bar{z}^{n-1}, \quad [z^{n},\bar{z}] = 2nHz^{n-1},  \nonumber \\ 
&& [z,z^{-n}] = 2nH|z|^{-2n}\bar{z}^{n-1}, \quad [z^{n},z^{-1}] = 2nH|z|^{-2}z^{n-1},  \quad 
( \forall n \in {\bf N} ).   
\end{eqnarray}
The conformal ( Virasoro ) algebra of the noncommutative $z$-plane 
will be obtained by using these commutators. 
We can mention that, the commutator $[z,\bar{z}]\ne 0$ introduces a ( modification of ) 
topology in complex manifolds. 


For example, limiting values of the following star products become
\begin{eqnarray}
\lim_{z\to 0}z\star\bar{z} &=& +H, \quad \lim_{z\to 0}\bar{z}\star z = -H,  \quad
\lim_{z\to 0}F(z\star \bar{z}) = F(+H), \quad \lim_{z\to 0}F(\bar{z}\star z) = F(-H),   
\end{eqnarray}
namely, they do not coincide at $H\ne 0$.
If $\bar{H}=-H$, $\lim_{z\to 0}\bar{z}\star z=\lim_{z\to 0}\overline{z\star\bar{z}}$.
A polynomial which has $n$ ( $\in{\bf N}$ ) zero points becomes
\begin{eqnarray}
P(z) &=& a_{0} + a_{1}z + a_{2}z^{2} + \cdots + a_{n}z^{n} = a_{n}(z-\alpha_{1})(z-\alpha_{2})\cdots(z-\alpha_{n}).
\end{eqnarray}
An entire function expanded around the origin of the $z$-plane will be given as
\begin{eqnarray}
f(z) &=& a_{0}+a_{1}z+a_{2}z^{2}+\cdots = \sum^{\infty}_{n=0}\frac{z^{n}}{n!}\frac{d^{n}f(z)}{dz^{n}}|_{z=0},  \\
\frac{d}{dz}z^{n} &=& nz^{n-1}, \quad \frac{d^{j}}{dz^{j}}z^{n} = \frac{n!}{(n-j)!}z^{n-j}, \quad 
( \forall n\ge j \in {\bf N} ),
\end{eqnarray}
where, a derivative has been defined as the ordinary case.
If the center of a series of an entire function is given at a point $\zeta\in\Omega$,
\begin{eqnarray}
g(z) &=& b_{0}+b_{1}(z-\zeta)+b_{2}(z-\zeta)^{2}+\cdots = \sum^{\infty}_{n=0}\frac{(z-\zeta)^{n}}{n!}\frac{d^{n}g(z)}{dz^{n}}|_{z=\zeta},  \\
\frac{d}{dz}(z-\zeta)^{n} &=& n(z-\zeta)^{n-1}, \quad 
\frac{d^{j}}{dz^{j}}(z-\zeta)^{n} = \frac{n!}{(n-j)!}(z-\zeta)^{n-j},  \quad ( \forall n\ge j \in {\bf N} ).
\end{eqnarray}
The exponential function $f(z) = \exp(z)$ is given as an absolutely convergent series,
infinitely differentiable, and satisfies the following relations:
\begin{eqnarray}
e^{z} &=& 1 + z + \frac{1}{2!}z^{2} + \cdots, \quad e^{\bar{z}} = \overline{e^{z}}, \quad
e^{z}e^{z} = e^{2z}, \nonumber \\
e^{z+\bar{z}} &=& e^{-H}e^{z}e^{\bar{z}} = e^{H}e^{\bar{z}}e^{z}, \quad 
e^{z+z^{-1}} = e^{-H|z|^{-2}}e^{z}e^{z^{-1}} = e^{H|z|^{-2}}e^{z^{-1}}e^{z},  \quad
e^{z^{-1}} = 1 + \frac{1}{z} + \frac{1}{2!z^{2}} + \cdots.
\end{eqnarray}
The analytic continuations of real functions $e^{\sigma_{1}+\sigma^{-1}_{1}}$ and $e^{2\sigma_{1}}$ 
to the complex $z$-plane are not unique, because 
$e^{\sigma_{1}+\sigma^{-1}_{1}}=e^{\sigma_{1}}e^{\sigma^{-1}_{1}}=e^{\sigma^{-1}_{1}}e^{\sigma_{1}}$, 
and $e^{2\sigma_{1}}=e^{\sigma_{1}+\sigma_{1}} = e^{\sigma_{1}}e^{\sigma_{1}}$ 
( $\sigma_{1}=\Re(z)$ ).
The exponential function $z\to e^{z}$ induces the mapping $\varphi:{\bf C}\to{\bf C}-\{0\}$,
and ${\bf C}$ is the universal covering surface of ${\bf C}-\{0\}$ also in the noncommutative case.
Transcendental functions $\cos z$ and $\sin z$ also be defined in the same way with the ordinary case.
The logarithmic function is determined by 
$\ln (1+z)= 2k\pi i + \sum^{\infty}_{n=1}(-1)^{n-1}z^{n}/n$ 
( $|z|<1$, $\forall k\in {\bf Z}$ ).
The geometric series is introduced, and the consideration for the summation formula
by the manipulations given below shows an exotic properties:
\begin{eqnarray}
S^{g}_{\infty} &=& 1 + z + z^{2} + \cdots,   \nonumber \\
S^{g}_{n} &=& \sum^{n}_{n=0}z^{n} = 1 + z + z^{2} + \cdots + z^{n-1}, \nonumber  \\
(1-z)S^{g}_{n} &=& S^{g}_{n}(1-z) = 1 - z^{n}.  
\end{eqnarray}
Because $(1-z)^{-1}(1-z)\ne(1-z)(1-z)^{-1}\ne 1$,
$S^{g}_{n}$ cannot be obtained in the forms as
$S^{g}_{n}=(1-z)^{-1}(1-z^{n})$ or  $S^{g}_{n}=(1-z^{n})(1-z)^{-1}$.
Poles and singularities are expressed by a Laurent expansion:
\begin{eqnarray}
f(z) &=& \cdots+\frac{a_{-2}}{z^{2}}+\frac{a_{-1}}{z}+a_{0}+a_{1}z+a_{2}z^{2}+\cdots  \nonumber \\
&=& \cdots + a_{-2}|z|^{-4}\bar{z}^{2} + a_{-1}|z|^{-2}\bar{z} + a_{0} + a_{1}z + a_{2}z^{2} + \cdots = \sum^{\infty}_{n=1}a_{-n}z^{-n} + \sum^{\infty}_{n=0}a_{n}z^{n}.
\end{eqnarray}
Here, we have presented the case of the essential singularity at the origin $z=0$.
Then the star product of $f(z)$ and $g(z)$ under taking forms of Laurent expansions becomes
\begin{eqnarray}
f(z)\star g(z) &=& 
\Bigl(\sum^{\infty}_{n=1}a_{-n}z^{-n}\Bigr)\star\Bigl(\sum^{\infty}_{n'=1}b_{-n'}z^{-n'}\Bigr) 
+ \Bigl(\sum^{\infty}_{n=0}a_{n}z^{n}\Bigr)\star\Bigl(\sum^{\infty}_{n'=0}b_{n'}z^{n'}\Bigr)   \nonumber \\
& & +\Bigl(\sum^{\infty}_{n=1}a_{-n}z^{-n}\Bigr)\star\Bigl(\sum^{\infty}_{n'=0}b_{n'}z^{n'}\Bigr)
+ \Bigl(\sum^{\infty}_{n=0}a_{n}z^{n}\Bigr)\star\Bigl(\sum^{\infty}_{n'=1}b_{-n'}z^{-n'}\Bigr)   \nonumber \\
&=& f(z)g(z) \ne g(z)f(z) = g(z)\star f(z).
\end{eqnarray}
In fact, if holomorphic functions $f(z)$ and $g(z)$ have singularities at $z=0$, 
following equation is satisfied:
\begin{eqnarray}
f(z)\star g(z) &=& f(z)g(z) \ne g(z)f(z) = g(z)\star f(z).
\end{eqnarray}
If $f(z)$ and $g(z)$ are entire functions ( no singularity ) of $z$, 
they commute and star products of them are identically the same with ordinary point products:
\begin{eqnarray}
f(z)\star g(z) &=& g(z)\star f(z) = f(z)g(z) = g(z)f(z),   \nonumber \\
f(z)\star g(z)\star h(z) &=& \bigl( f(z)\star g(z) \bigr) \star h(z) = f(z)\star \bigl( g(z)\star h(z) \bigr). 
\end{eqnarray}
Hence, the associativity is satisfied in the star products of entire functions $f(z)$, $g(z)$, $\cdots$.
In this case, the Poisson and symplectic structures are not introduced in the star products.
The effect of noncommutativity may arise through taking a star product of a holomorphic function $f(z)$ 
and an antiholomorphic function $\bar{g}(\bar{z})$,
and the star product of them is not holomorphic also in the noncommutative case.


If we repeat the discussion on the condition of holomorphicity of a function, 
it is the same also in our case, namely, the fulfillment of the Cauchy-Riemann equations: 
$\partial_{\bar{z}}f=0$.
The harmonic function condition $\partial_{z}\partial_{\bar{z}}f=0$ is satisfied in that case.
It is also the case that, a holomorphic mapping with $f'(z)\ne 0$ 
becomes conformal on a noncommutative Riemann surface.
$1/(z-\alpha)$ is holomorphic except $z=\alpha$, has a order 1 pole at $z=\alpha$, 
and a formal expansion will be given by
\begin{eqnarray}
\frac{1}{z-\alpha} &\equiv& (z-\alpha)^{-1} 
= z^{-1} + \alpha z^{-2} + \alpha^{2}z^{-3} + \cdots = \sum^{\infty}_{n=1}\alpha^{n-1}z^{-n}, \quad 
( \alpha \in \Omega, \alpha^{0}=1 ).
\end{eqnarray}
The expression of the expansion has been defined in the region where the series converges: 
$|\alpha|/|z|<1$ with $|z|\equiv\sqrt{\sigma^{2}_{1}+\sigma^{2}_{2}}$.
Therefore, one finds
\begin{eqnarray}
&& \Bigl[z-\beta,\frac{1}{z-\alpha}\Bigr] = \sum^{\infty}_{n=1}\alpha^{n-1}[z,z^{-n}] 
= 2H\sum^{\infty}_{n=1}n|z|^{-2n}(\alpha\bar{z})^{n-1}, \quad ( +\infty > |z| > |\alpha|, \alpha\ne 0 ).
\end{eqnarray}
In the discussion given above, we have assumed $[z,\alpha]=[z,\beta]=0$.


Because the noncommutative $z$-plane still keeps the Hausdorff topological character,
the idea of Weierstrass-Weyl analytic continuation on noncommutative Riemann surfaces 
obtains interesting phenomena.
As the discussion of Weyl on ${\it analytisches}$ ${\it Gebilde}$~[15], 
we generalize the form of power series as
\begin{eqnarray}
\varphi: z = P(t) \longrightarrow w = Q(t), \quad (z,w,t\in {\bf C}),
\end{eqnarray} 
where, $t$ is a complex parameter, and both $P(t)$ and $Q(t)$ include ( at most ) 
finite orders of negative powers of $t$ ${\it in}$ ${\it general}$.
This is just an example of the definition of a complex manifold
if we supplement the differentiable condition for the mapping $\varphi$ or $P^{-1}\circ Q$.
In that case, the commutator will be expressed as follows:
\begin{eqnarray}
& & [z,\bar{z}] = [P(t),\bar{P}(\bar{t})] = 2H.
\end{eqnarray}
The viewpoint of the ${\it analytisches}$ ${\it Gebilde}$ ${\bf G}$ provides a new interpretation on the
noncommutativity. For example, if $z$ is expanded by new complex parameters $t$ and $\tau$ 
in the following forms of entire functions, assumed as they are convergent series,
\begin{eqnarray}
z &=& a_{0} + a_{1}t + a_{2}t^{2} + \cdots  = \sum_{n=0} a_{n}t^{n}  \nonumber \\
&=& b_{0} + b_{1}\tau + b_{2}\tau^{2} + \cdots = \sum_{m=0}b_{m}\tau^{m}, 
\end{eqnarray}
we recognize that, the introduction of the commutator $[z,\bar{z}]=2H$ 
should be interpreted as a choice for the representation of the noncommutativity,
and it is based on the infinite-dimensional linear algebra ( linear group ) 
of the ${\it analytisches}$ ${\it Gebilde}$ ${\bf G}$:
\begin{eqnarray}
2H &=& [z,\bar{z}] = [Uz'U^{-1},U\bar{z}'U^{-1}] 
= \sum_{n,n'}[a_{n}t^{n},\bar{a}_{n'}\bar{t}^{n'}] = \sum_{m,m'}[b_{m}\tau^{m},\bar{b}_{m'}\bar{\tau}^{m'}],
\end{eqnarray}
where, $U$ is a formally defined ( finite or infinite dimensional ) ${\it unitary}$ matrix 
for changing representations.
Equation (49) says that, $[t^{n},\bar{t}^{n'}]\ne 0$ in general.
Hence $[z,\bar{z}]=2H$ is the simplest expression for the noncommutativity.
The discussion just has shown the existence of an equivalence class, related to the concepts of germ and sheaf.
We have arrived at a sheaf, namely a Riemann surface of a global analytic function.
Moreover, in a path integration of a string theory action, 
a diffeomorphic mapping $\varphi:\Sigma\to\Sigma$, $z\to z'(z)$ ( $\Sigma$; a Riemann surface ) 
corresponds to a gauge degree of freedom, 
and hence the commutator $[z,\bar{z}]=2H$ may fix the gauge in a string theory. 
The star product itself is not invariant under the transformation (46),
and when $z=P(t)$ has its inverse $t=P^{-1}(z)$,
\begin{eqnarray}
&& f(z,\bar{z})\star g(z,\bar{z}) = f(P(t),\bar{P}(\bar{t}))\star g(P(t),\bar{P}(\bar{t}))   \nonumber \\
&& \quad = f(P(t),\bar{P}(\bar{t}))e^{H[
\overleftarrow{\partial}_{t}(\partial_{z}P^{-1}(z(t)))
(\partial_{\bar{z}}\bar{P}^{-1}(\bar{z}(\bar{t})))\overrightarrow{\partial}_{\bar{t}}
-
\overleftarrow{\partial}_{\bar{t}}(\partial_{\bar{z}}\bar{P}^{-1}(\bar{z}(\bar{t})))
(\partial_{z}P^{-1}(z(t)))\overrightarrow{\partial}_{t}
]}g(P(t),\bar{P}(\bar{t})).
\end{eqnarray}
The analytic continuation provides an interesting problem for us, 
because the center of a power series expansion will move step by step in a sequence of the procedure
of continuation: 
\begin{eqnarray}
f(z) &=& c_{0} + c_{1}z + c_{2}z^{2} + \cdots   \nonumber \\
&=& \tilde{c}_{0} + \tilde{c}_{1}(z-\alpha) + \tilde{c}_{2}(z-\alpha)^{2} + \cdots  \nonumber \\
&=& \breve{c}_{0} + \breve{c}_{1}(z-\beta) + \breve{c}_{2}(z-\beta)^{2} + \cdots,   
\end{eqnarray}
and it should be examined whether the method can be used in noncommutative Riemann surfaces or not,
sometimes determined by functions they have singularities,
because the noncommutativity (13) fixes the origin of the $z$-plane.   
If an analytic continuation exists, the monodromy theorem
of two homotopic curves trivially realizes by the procedure of continuation.


Various relations will be derived from the commutator $[z,\bar{z}]=2H$:
\begin{eqnarray}
&& z\star z = z^{2}, \quad \bar{z}\star\bar{z} = \bar{z}^{2}, \quad 
z\star\bar{z} = z\bar{z} + H, \quad \bar{z}\star z = \bar{z}z - H,   \nonumber  \\
&& \overline{z\star\bar{z}} = \overline{z\bar{z} + H} = \bar{z}z + \bar{H} = \bar{z}\star z + H + \bar{H},   \nonumber \\
&& z\star\bar{z} + \bar{z}\star z = z\bar{z} + \bar{z}z, \quad z\star\bar{z} - \bar{z}\star z = 2H,   \nonumber \\
&& (z\star\bar{z})\star z = z\star(\bar{z}\star z) = z\bar{z}z,  \quad
(\bar{z}\star z)\star\bar{z} = \bar{z}\star(z\star\bar{z}) = \bar{z}z\bar{z},  \nonumber \\
&& (z\star\bar{z})\star(z\star\bar{z}) = ((z\star\bar{z})\star z)\star\bar{z} = z\star(\bar{z} \star (z\star\bar{z})) = z\bar{z}z\bar{z} + H(z\bar{z}+\bar{z}z),
\end{eqnarray}
and
\begin{eqnarray}
z^{l}\star\bar{z}^{m} 
&=& z^{l}\bar{z}^{m} + Hlmz^{l-1}\bar{z}^{m-1} + \frac{H^{2}}{2!}l(l-1)m(m-1)z^{l-2}\bar{z}^{m-2} 
+ \cdots + H^{m}\frac{l!}{m!}z^{l-m} \quad ( l > m )   \nonumber \\
&=& z^{l}\bar{z}^{m} + Hlmz^{l-1}\bar{z}^{m-1} + \frac{H^{2}}{2!}l(l-1)m(m-1)z^{l-2}\bar{z}^{m-2} 
+ \cdots + H^{l}\frac{m!}{l!}\bar{z}^{m-l} \quad ( l < m ),  \nonumber \\
& &  ( \forall l,m \in {\bf N} ).
\end{eqnarray}
We have found that, the symmetry under the operation of complex conjugation is broken in our theory
${\it in}$ ${\it general}$.
From $[z,\bar{z}]=2H$, one finds it is favorable if $H$ is a pure imaginary number, $\bar{H}=-H$.
( The conjugation $\rho: [z,\bar{z}]\to [\bar{z},z]$ shows that 
$\rho: z\bar{z}-\bar{z}z=2H\to\bar{z}z-z\bar{z}=2\bar{H}$. )
In that case, $\overline{z\star\bar{z}}=\bar{z}\star z$ is satisfied.
Now, one finds $(z\pm\bar{z})(z\mp\bar{z}) = z^{2}-\bar{z}^{2}\mp 2H = 4i\Re (z)\Im (z)$.
Hence,
\begin{eqnarray}
H &=& \frac{1}{2}(z^{2}-\bar{z}^{2}) - 2i\Re(z)\Im(z) = \frac{1}{2}(\bar{z}^{2}-z^{2}) + 2i\Im(z)\Re(z), 
\end{eqnarray}
i.e., $H$ has been expressed by $z$ and $\bar{z}$.
It is possible for us to confirm that, this expression for $H$ given by $z$ and $\bar{z}$ vanishes ( $H=0$ )
when $z$ and $\bar{z}$ commute with each other.
Several algebraic relations of fractions are found to be 
\begin{eqnarray}
\frac{1}{z} &\equiv& z^{-1}, \quad \frac{1}{\bar{z}} \equiv \bar{z}^{-1},  \nonumber \\
z\star\frac{1}{z} &\ne& \frac{z\star\bar{z}}{|z|^{2}}, \quad 
z\star\frac{1}{\bar{z}} \ne \frac{z\star z}{|z|^{2}}, \quad 
\bar{z}\star\frac{1}{z} \ne \frac{\bar{z}\star\bar{z}}{|z|^{2}}, \quad 
\bar{z}\star\frac{1}{\bar{z}} \ne \frac{\bar{z}\star z}{|z|^{2}},   \nonumber \\
z\star\frac{1}{z} &=& zz^{-1},  \quad
z\star\frac{1}{\bar{z}} = z\bar{z}^{-1}-H\bar{z}^{-2}, \quad
\bar{z}\star\frac{1}{z} = \bar{z}z^{-1}+Hz^{-2}, \quad
\bar{z}\star\frac{1}{\bar{z}} = \bar{z}\bar{z}^{-1}, \nonumber \\
\frac{1}{z}\star z &=& z^{-1}z, \quad 
\frac{1}{z}\star\bar{z} = z^{-1}\bar{z}-Hz^{-2}, \quad
\frac{1}{\bar{z}}\star z = \bar{z}^{-1}z+H\bar{z}^{-2},  \quad 
\frac{1}{\bar{z}}\star\bar{z} = \bar{z}^{-1}\bar{z}, \nonumber \\
\frac{1}{z}\star\frac{1}{z} &=& z^{-2}, \quad 
\frac{1}{z}\star\frac{1}{\bar{z}} = z^{-1}\bar{z}^{-1}+Hz^{-2}\bar{z}^{-2}+2H^{2}z^{-3}\bar{z}^{-3}+\cdots, \nonumber \\ 
\frac{1}{\bar{z}}\star\frac{1}{\bar{z}} &=& \bar{z}^{-2},  \quad
\frac{1}{\bar{z}}\star\frac{1}{z} = \bar{z}^{-1}z^{-1}-H\bar{z}^{-2}z^{-2}+2H^{2}\bar{z}^{-3}z^{-3}+\cdots,
\end{eqnarray}
and
\begin{eqnarray}
z^{l}\star\bar{z}^{-m} &=& z^{l}\bar{z}^{-m} + (-H)lmz^{l-1}\bar{z}^{-m-1}
+ \frac{(-H)^{2}}{2!}l(l-1)m(m+1)z^{l-2}\bar{z}^{-m-2}   \nonumber \\
& &  + \cdots
+ (-H)^{l}m(m-1)\cdots(m+l-1)\bar{z}^{-m-l}, \quad ( \forall l,m \in {\bf N} ).
\end{eqnarray}
Because the star product is defined by operations of $\partial_{z}$ and $\partial_{\bar{z}}$,
we have to carefully handle functions of $z$ and $\bar{z}$.
It is possible for us to manipulate the star products by converting the expression of 
$z^{-1}$ and $\bar{z}^{-1}$ given in terms of $\sigma_{1}$ and $\sigma_{2}$. 
Sometimes this method is useful for doing confirmations.
The radical ( square ) roots are defined in the ordinary way, and they obey the following relations:
\begin{eqnarray}
w &\equiv& z^{1/n},  \quad  w^{n} = z,    \quad
(z^{1/2})^{2} = z, \quad z^{-1/2} = (z^{-1})^{1/2} = \frac{\bar{z}^{1/2}}{|z|},   \nonumber \\
z &=& z^{1/n}z^{1/n}\cdots z^{1/n} = (z^{1/n})^{n},  \quad
\bar{z} = \bar{z}^{1/n}\bar{z}^{1/n}\cdots \bar{z}^{1/n} = (\bar{z}^{1/n})^{n}, \quad
z^{1/n}\bar{z}^{1/n}\ne (z\bar{z})^{1/n}.
\end{eqnarray}
Hence,
\begin{eqnarray}
z^{1/2}\star z^{-1/2} &\ne& \frac{z^{1/2}\star \bar{z}^{1/2}}{|z|}, 
\quad z^{1/2}\star\bar{z}^{1/2} \ne (z\star\bar{z})^{1/2},  \nonumber \\
z\star z^{1/2} &=& z^{1/2}\star z = z^{3/2}, \quad
z\star\bar{z}^{1/2} = z\bar{z}^{1/2} + \frac{H}{2}\bar{z}^{-1/2}, \quad
\bar{z}^{1/2}\star z = \bar{z}^{1/2}z -\frac{H}{2}\bar{z}^{-1/2},  \nonumber \\
\bar{z}\star\bar{z}^{1/2} &=& \bar{z}^{1/2}\star\bar{z} = \bar{z}^{3/2}, \quad
\bar{z}\star z^{1/2} = \bar{z}z^{1/2} -\frac{H}{2}z^{-1/2}, \quad
z^{1/2}\star\bar{z} = z^{1/2}\bar{z} + \frac{H}{2}z^{-1/2},   \nonumber \\
z^{1/2}\star z^{1/2} &=& z^{1/2}z^{1/2} = z, \quad \bar{z}^{1/2}\star\bar{z}^{1/2} = \bar{z}^{1/2}\bar{z}^{1/2} = \bar{z},   \nonumber \\
z^{1/2}\star\bar{z}^{1/2} &=& z^{1/2}\bar{z}^{1/2} + \frac{H}{4}z^{-1/2}\bar{z}^{-1/2},  + \cdots, \quad
\bar{z}^{1/2}\star z^{1/2} = \bar{z}^{1/2}z^{1/2} - \frac{H}{4}\bar{z}^{-1/2}z^{-1/2},  + \cdots,    \nonumber \\
z^{1/2}\star z^{-1/2} &=& z^{1/2}z^{-1/2}, \quad
z^{1/2}\star \bar{z}^{-1/2} = z^{1/2}\bar{z}^{-1/2} -\frac{H}{4}z^{-1/2}\bar{z}^{-3/2} + \cdots,  \nonumber \\
z^{-1/2}\star z^{1/2} &=& z^{-1/2}z^{1/2}, \quad
\bar{z}^{-1/2}\star z^{1/2} = \bar{z}^{-1/2}z^{1/2} + \frac{H}{4}\bar{z}^{-3/2}z^{-1/2} + \cdots.
\end{eqnarray}
It should be noticed that, both $z^{1/2}\star\bar{z}^{1/2}$ and $\bar{z}^{1/2}\star z^{1/2}$  
are infinite-order series.


Next, we consider the following commutators:
\begin{eqnarray}
&& [z^{2},\bar{z}] = 4Hz, \quad [z,\bar{z}^{2}] = 4H\bar{z} , \nonumber \\
&& [z^{2},\bar{z}^{2}] = 8Hz\bar{z}-8H^{2}, \quad [z^{3},\bar{z}] = 6Hz^{2}, \quad  
[z,\bar{z}^{3}] = 6H\bar{z}^{2}, \nonumber \\
&& [z^{3},\bar{z}^{2}] =  12Hz^{2}\bar{z}-24H^{2}z, \quad 
[z^{2},\bar{z}^{3}] = 12Hz\bar{z}^{2} - 24H^{2}\bar{z}, \nonumber \\
&& [z^{3},\bar{z}^{3}] = 18Hz^{2}\bar{z}^{2} - 88H^{2}z\bar{z} +48H^{3},
\end{eqnarray}
and,
\begin{eqnarray}
&& [z,\bar{z}^{n}] = 2nH\bar{z}^{n-1} 
= 2H\frac{d\bar{z}^{n}}{d\bar{z}} = 2H\Bigl[\frac{d}{d\bar{z}},\bar{z}^{n}\Bigr],  \nonumber \\  
&& [z^{n},\bar{z}] = 2nHz^{n-1} 
= 2H\frac{dz^{n}}{dz} = -2H\Bigl[z^{n},\frac{d}{dz}\Bigr],   \nonumber \\
&& [z^{2},\bar{z}^{n}] = z[z,\bar{z}^{n}] + [z,\bar{z}^{n}]z,  \quad  
   [z^{n},\bar{z}^{2}] = \bar{z}[z^{n},\bar{z}] + [z^{n},\bar{z}]\bar{z},   \nonumber \\
&& [z^{3},\bar{z}^{n}] = z^{2}[z,\bar{z}^{n}] + z[z,\bar{z}^{n}]z + [z,\bar{z}^{n}]z^{2},  \quad 
[z^{n},\bar{z}^{3}] = \bar{z}^{2}[z^{n},\bar{z}] + \bar{z}[z^{n},\bar{z}]\bar{z} + [z^{n},\bar{z}]\bar{z}^{2},  \nonumber \\
&& [z^{l},\bar{z}^{n}] 
= 2nH\sum^{l}_{i=1}z^{i-1}\bar{z}^{n-1}z^{l-i} = 2lH\sum^{n}_{j=1}\bar{z}^{n-j}z^{l-1}\bar{z}^{j-1}, \quad
( \forall l,n\in{\bf N} ).
\end{eqnarray}
Moreover,
\begin{eqnarray}
&& \overbrace{[z,[z,\cdots,[z}^{m},\bar{z}^{n}]]\cdots] = (2H)^{m}\frac{n!}{m!}\bar{z}^{n-m} 
= (2H)^{m}\frac{d^{m}\bar{z}^{n}}{d\bar{z}^{m}}
= (2H)^{m}\overbrace{\Bigl[\frac{d}{d\bar{z}},\Bigl[\frac{d}{d\bar{z}},\cdots,\Bigl[\frac{d}{d\bar{z}}}^{m},\bar{z}^{n}\Bigr]\Bigr]\cdots\Bigr],  \nonumber  \\
&& [\cdots [[z^{n},\overbrace{\bar{z}],\cdots,\bar{z}],\bar{z}]}^{m} = (2H)^{m}\frac{n!}{m!}z^{n-m}, 
= (2H)^{m}\frac{d^{m}z^{n}}{dz^{m}}
= (-2H)^{m}\Bigl[\cdots \Bigl[\Bigl[z^{n},\overbrace{\frac{d}{dz}\Bigr],\cdots,\frac{d}{dz}\Bigr],\frac{d}{dz}\Bigr]}^{m},  \nonumber  \\   
&& ( \forall l, m, n \in {\bf N}, n > m ).  
\end{eqnarray}
Here, we have converted the expressions of the commutators by using the derivatives 
$\frac{d}{dz}$ and $\frac{d}{d\bar{z}}$, because $\bar{z}$ and $2H\frac{d}{dz}$ 
( $z$ and $2H\frac{d}{d\bar{z}}$ ) are equivalent inside the commutators
$[z^{n},\bar{z}]$ ( $[z,\bar{z}^{n}]$ ).
$2H\partial_{z}$ ( $\in TM^{+}=\bigcup_{p\in M^{+}}T_{p}M^{+}$ ) 
and $2H\partial_{\bar{z}}$ ( $\in TM^{-}=\bigcup_{p\in M^{-}}T_{p}M^{-}$ ) are regarded as tangent vectors,
where $TM^{+}$ and $TM^{-}$ are a holomorphic and an antiholomorphic tangent vector bundles, respectively
( $M^{+}$ and $M^{-}$ are base manifolds of the vector bundles ).
Hence, we can summarize these results into the following similarity transformations,
\begin{eqnarray}
(\bar{z}')^{n} &\equiv& e^{z}\bar{z}^{n}e^{-z} 
= \bar{z}^{n} + [z,\bar{z}^{n}] + \frac{1}{2!}[z,[z,\bar{z}^{n}]] + \cdots \nonumber \\
&=& \bar{z}^{n} + 2Hn\bar{z}^{n-1} + \frac{n(n-1)}{2!}(2H)^{2}\bar{z}^{n-2} + \cdots  \nonumber \\
&=& \Bigl( 1+2H\frac{d}{d\bar{z}}+\frac{(2H)^{2}}{2!}\frac{d^{2}}{d\bar{z}^{2}}+\cdots \Bigr)\bar{z}^{n} 
= e^{2H\frac{d}{d\bar{z}}}\bar{z}^{n},  \nonumber  \\ 
(z')^{n} &\equiv& e^{-\bar{z}}z^{n}e^{\bar{z}} 
= z^{n} + [z^{n},\bar{z}] + \frac{1}{2!}[[z^{n},\bar{z}],\bar{z}] + \cdots \nonumber \\
&=& z^{n} + 2Hnz^{n-1} + \frac{n(n-1)}{2!}(2H)^{2}z^{n-2} + \cdots   \nonumber \\
&=& \Bigl( 1+2H\frac{d}{dz}+\frac{(2H)^{2}}{2!}\frac{d^{2}}{dz^{2}}+\cdots \Bigr)z^{n}
= e^{2H\frac{d}{dz}}z^{n},  \quad ( \forall n \in {\bf N} ). 
\end{eqnarray}
The specific case $n=1$ of the above results corresponds to a canonical transformation
satisfying $\{z',\bar{z}'\}^{P.B.}_{z,\bar{z}}=1$.
These transformations show a pseudo Lie-group structure in our theory.
The final expressions 
$(z')^{n}=\exp[2H\partial_{z}]z^{n}$ and $(\bar{z}')^{n}=\exp[2H\partial_{\bar{z}}]\bar{z}^{n}$
have the form of exponential mappings, and the norm of the operators will be estimated by
\begin{eqnarray}
\|2H\partial_{z}\| &\equiv& \sup_{0\ne z\in D}\frac{\|2H\partial_{z}(z)^{n}\|}{\|z^{n}\|}
\sim \sup_{0\ne z\in D}|2nH||z|^{-1}, \nonumber \\
\|2H\partial_{\bar{z}}\| &\equiv& \sup_{0\ne \bar{z}\in \bar{D}}\frac{\|2H\partial_{\bar{z}}(\bar{z})^{n}\|}{\|\bar{z}^{n}\|}
\sim \sup_{0\ne \bar{z}\in \bar{D}} |2nH||z|^{-1},   \nonumber \\
& & ( {\rm with}\, \|z^{n}\|\sim |z|^{n} ).
\end{eqnarray}
This estimation says that both of the convergence radii of the exponential mappings are finite
under the case of a bounded Riemann surface.
Later, we will discuss several definitions of norm in the context of functional analysis.
Furthermore, one finds the following relations 
for transformations of combinations of partial derivatives $\partial_{z}$, $\partial_{\bar{z}}$ 
and $z^{n}$, $\bar{z}^{n}$:
\begin{eqnarray}
e^{\alpha z}(\partial_{z}+\beta z^{n})e^{-\alpha z} &=& \partial_{z}-\alpha+\beta z^{n},  \nonumber  \\
e^{\alpha z}(\partial_{\bar{z}}+\bar{\beta}\bar{z}^{n})e^{-\alpha z} 
&=& \partial_{\bar{z}}+\bar{\beta}\bigl[ e^{2H\alpha\partial_{\bar{z}}}\bar{z}^{n}\bigr],   \nonumber \\
e^{-\bar{\alpha}\bar{z}}(\partial_{z}+\beta z^{n})e^{\bar{\alpha}\bar{z}} 
&=& \partial_{z} + \beta\bigl[ e^{2H\bar{\alpha}\partial_{z}}z^{n}\bigr],   \nonumber \\
e^{-\bar{\alpha}\bar{z}}(\partial_{\bar{z}}+\bar{\beta}\bar{z}^{n})e^{\bar{\alpha}\bar{z}} 
&=& \partial_{\bar{z}}+\bar{\alpha}+\bar{\beta}\bar{z}^{n}, \quad ( \alpha, \beta \in {\bf C} ).
\end{eqnarray}
Because the functions $z$ and $\bar{z}$ are odd under $z\to -z$ and $\bar{z}\to -\bar{z}$,
these transformations have the parity-odd characters if $n$ is an odd number. 
For example, a more general case, the similarity transformation of an entire function 
$f(z) = a_{0}+ a_{1}z + a_{2}z^{2} + \cdots$ and its complex conjugate become
\begin{eqnarray}
F_{+}(H) &\equiv& e^{-\bar{z}}f(z)e^{\bar{z}} = \exp\Bigl[ 2H\frac{d}{dz} \Bigr]f(z),   \quad
\delta_{z} f(z) = 2H\frac{d}{dz}f(z),    \nonumber \\ 
F_{-}(H) &\equiv& e^{z}\bar{f}(\bar{z})e^{-z} = \exp\Bigl[ 2H\frac{d}{d\bar{z}} \Bigr]\bar{f}(\bar{z}),  \quad
\delta_{\bar{z}} \bar{f}(\bar{z}) = 2H\frac{d}{d\bar{z}}\bar{f}(\bar{z}),
\end{eqnarray}
where, $F_{+}$ ( $F_{-}$ ) is a $C^{\infty}$-function of $H$, and (anti)holomorphic on $z$ ( $\bar{z}$ ).
The elements of the group are defined through mappings $\alpha\to\exp(\alpha z)$ and 
$\bar{\alpha}\to\exp(-\bar{\alpha}\bar{z})$, 
\begin{eqnarray}
&& \{ G(\alpha)\equiv e^{\alpha z}|\alpha\in{\bf C}\}, \quad
\{ \bar{G}(\bar{\alpha})\equiv e^{-\bar{\alpha}\bar{z}}|\bar{\alpha}\in{\bf C}\}, 
\end{eqnarray}
and they have the structure(s) of a one-parametric transformation group with the bases $z$ and $\bar{z}$.
$G(\alpha)$ ( $\bar{G}(\bar{\alpha})$ ) is a $C^{\infty}$-function of $\alpha$ ( $\bar{\alpha}$ ),
and obey the following relations:
\begin{eqnarray}
&& G(\alpha)^{-1} = e^{-\alpha z}, \quad \bar{G}(\bar{\alpha})^{-1} = e^{\bar{\alpha}\bar{z}},  \nonumber \\
&& \overline{G(\alpha)} = \bar{G}(-\bar{\alpha}), \quad
\overline{G(\alpha)^{-1}} = \bar{G}(-\bar{\alpha})^{-1}, \nonumber \\
&& G(\alpha)G(\beta) = G(\alpha+\beta), \quad 
\bar{G}(\bar{\alpha})\bar{G}(\bar{\beta}) = \bar{G}(\bar{\alpha}+\bar{\beta}),  \nonumber \\
&& G(\beta)G(\alpha)G(\beta)^{-1} = G(\alpha), \quad
G(\beta)\bar{G}(\bar{\alpha})G(\beta)^{-1} = e^{-2H\bar{\alpha}\beta}\bar{G}(\bar{\alpha}),  \nonumber \\ 
&& \bar{G}(\bar{\beta})G(\alpha)\bar{G}(\bar{\beta})^{-1} = e^{2H\alpha\bar{\beta}}\bar{G}(\bar{\alpha}), \quad
\bar{G}(\bar{\beta})\bar{G}(\bar{\alpha})\bar{G}(\bar{\beta})^{-1} = \bar{G}(\bar{\alpha}), \quad
\alpha, \beta \in {\bf C}.
\end{eqnarray}
Here we have observed that, 
both $G(\alpha)$ and $\bar{G}(\bar{\alpha})$ independently construct the Abelian groups ( module ).
Now, we examine the following functions:
\begin{eqnarray}
V_{+} &\equiv& \frac{d}{dH}F_{+}(H)\Big|_{H=0} = 2\frac{d}{dz}f(z) = \delta_{z}\frac{f(z)}{H},  \nonumber  \\
V_{-} &\equiv& \frac{d}{dH}F_{-}(H)\Big|_{H=0} = 2\frac{d}{d\bar{z}}\bar{f}(\bar{z}) 
= \delta_{\bar{z}}\frac{\bar{f}(\bar{z})}{H}.  
\end{eqnarray}
If we consider a variation problem of a functional ${\cal F}[F_{+},F_{-}]$ 
given in terms of $F_{+}$ and/or $F_{-}$, $V_{+}$ ( a holomorphic vector ) and $V_{-}$ 
( an antiholomorphic vector ) might be interpreted as variation vector fields, 
when the noncommutativity parameter $H$ acts as a variation parameter:
\begin{eqnarray}
\frac{\partial}{\partial H}{\cal F}[F_{+},F_{-}]\Big|_{H=0} &=& 
\frac{\delta{\cal F}}{\delta F_{+}}\frac{\partial F_{+}}{\partial H}\Big|_{H=0} +
\frac{\delta{\cal F}}{\delta F_{-}}\frac{\partial F_{-}}{\partial H}\Big|_{H=0} 
= \frac{\delta{\cal F}}{\delta F_{+}}V_{+} + \frac{\delta{\cal F}}{\delta F_{-}}V_{-}.
\end{eqnarray}
We consider the following eigenvalue problem:
\begin{eqnarray}
\hbar &\equiv& iH, \quad \Im(\hbar) = 0,   \nonumber \\
\frac{i}{2}\hbar V_{+} &=& i\hbar\frac{d}{dz}f(z) = \epsilon f(z), \quad
-\frac{i}{2}\hbar V_{-} = -i\hbar\frac{d}{d\bar{z}}\bar{f}(\bar{z}) = \bar{\epsilon}\bar{f}(\bar{z}).
\end{eqnarray}
Here, we have chosen $H$ as a pure imaginary number.
Therefore,
\begin{eqnarray} 
f(z) &=& C_{1}\exp\Bigl[ -\frac{i}{\hbar}\epsilon z \Bigr], \quad 
\bar{f}(\bar{z}) = C_{2}\exp\Bigl[ +\frac{i}{\hbar}\bar{\epsilon} \bar{z}\Bigr],
\end{eqnarray}
where, $C_{1}$ and $C_{2}$ are integration constants.
By using these solutions, one gets,
\begin{eqnarray}
{\cal F}[f(z),\bar{f}(\bar{z})] &\to& {\cal F}[F_{+},F_{-}] 
= {\cal F}[-2C_{1}\epsilon e^{-i\epsilon z/\hbar}, 2C_{2}\bar{\epsilon}e^{+i\bar{\epsilon}\bar{z}/\hbar}]. 
\end{eqnarray}


Now, we introduce the following definition of ${\it normal}$ ${\it ordering}$:
\begin{eqnarray}
f(z,\bar{z}) &=& z^{l}\bar{z}^{m}, \quad g(z,\bar{z}) = z^{n}\bar{z}^{j},   \nonumber \\
:f(z,\bar{z})g(z,\bar{z}): &=& :z^{l}\bar{z}^{m}z^{n}\bar{z}^{j}:   \nonumber \\
&=& z^{l+n}\bar{z}^{m+j} + H (c_{1}z^{l+n-1}\bar{z}^{m+j-1}) + H^{2}(c_{2}z^{l+n-2}\bar{z}^{m+j-2}) + \cdots,  \nonumber \\
& & ( \forall l, m, n, j \in {\bf N} ),
\end{eqnarray}
where, all of $z$ take positions of the left of all of $\bar{z}$.
$c_{1},c_{2},\cdots$ are coefficients determined by the numbers $l,m,n,j$.
In fact, $:f(z,\bar{z})g(z,\bar{z}):$ is a polynomial of $H$.
Because of the ambiguities coming from the noncommutativity,
we employ the normal ordering before taking a derivative of $z$.
Here, we should emphasize that this situation of the ambiguity is also the case
in the evaluation of the star product.
This is a quite important problem in the context of deformation quantization.
For example, we will define ( choose ) that 
a derivative $\frac{\overrightarrow{d}}{dz}$ will act as a "left-derivative" 
while $\frac{\overleftarrow{d}}{d\bar{z}}$ acts as a "right-derivative"
to a normal-ordered $z^{l}\bar{z}^{m}$ similar to the situations of Grassmann numbers: 
\begin{eqnarray}
\frac{\overrightarrow{d}}{dz}:z^{l}\bar{z}^{m}: &=& lz^{l-1}\bar{z}^{m},  \quad
\frac{\overrightarrow{d}}{dz}:z^{l}\bar{z}^{m}:\frac{\overleftarrow{d}}{d\bar{z}} = lmz^{l-1}\bar{z}^{m-1},  \nonumber \\
\frac{\overrightarrow{d}}{dz}:z^{l}\bar{z}^{m}z^{n}\bar{z}^{j}: &=& (l+n)z^{l+n-1}\bar{z}^{m+j}  \nonumber \\
& & + Hc_{1}(l+n-1)z^{l+n-2}\bar{z}^{m+j-1} + H^{2}c_{2}(l+n-2)z^{l+n-3}\bar{z}^{m+j-2} +\cdots.
\end{eqnarray}
The derivatives of normal-ordered commutators are found to be
\begin{eqnarray}
&& \frac{\overrightarrow{d}}{dz}:[z,\bar{z}]: = 0,  \nonumber \\
&& \frac{\overrightarrow{d}}{dz}:[z^{2},\bar{z}]: = 2:[z,\bar{z}]: = 4H, \quad
\frac{\overrightarrow{d}}{dz}:[z,\bar{z}^{2}]: = 0,  \quad
\frac{\overrightarrow{d}}{dz}:[z^{2},\bar{z}^{2}]: = 2:[z,\bar{z}^{2}]: = 8H\bar{z},  \nonumber \\
&& \frac{\overrightarrow{d}}{dz}:[z,\bar{z}^{3}]: = 0,  \quad
\frac{\overrightarrow{d}}{dz}:[z^{3},\bar{z}]: = 3:[z^{2},\bar{z}]: = 12Hz,  \nonumber \\
&& \frac{\overrightarrow{d}}{dz}:[z^{2},\bar{z}^{3}]: = 2:[z,\bar{z}^{3}]: = 12H\bar{z}^{2},  \quad
\frac{\overrightarrow{d}}{dz}:[z^{3},\bar{z}^{2}]: = 3:[z^{2},\bar{z}^{2}]: = 24Hz\bar{z}-24H^{2},  \nonumber \\
&& \frac{\overrightarrow{d}}{dz}:[z^{3},\bar{z}^{3}]: = 3:[z^{2},\bar{z}^{3}]: = 36Hz\bar{z}^{2}-88H^{2}\bar{z},  
\end{eqnarray}
and,
\begin{eqnarray}
&& :[z,\bar{z}]:\frac{\overleftarrow{d}}{d\bar{z}} = 0,  \nonumber \\
&& :[z^{2},\bar{z}]:\frac{\overleftarrow{d}}{d\bar{z}} = 0, \quad
:[z,\bar{z}^{2}]:\frac{\overleftarrow{d}}{d\bar{z}} = 2:[z,\bar{z}]: = 4H,  \quad
:[z^{2},\bar{z}^{2}]:\frac{\overleftarrow{d}}{d\bar{z}} = 2:[z^{2},\bar{z}]: = 8Hz,  \nonumber \\
&& :[z,\bar{z}^{3}]:\frac{\overleftarrow{d}}{d\bar{z}} = 3:[z,\bar{z}^{2}]: = 12H\bar{z},  \quad
:[z^{3},\bar{z}]:\frac{\overleftarrow{d}}{d\bar{z}} = 0,  \nonumber \\
&& :[z^{2},\bar{z}^{3}]:\frac{\overleftarrow{d}}{d\bar{z}} = 3:[z^{2},\bar{z}^{2}]: = 24Hz\bar{z}-24H^{2},  \quad
:[z^{3},\bar{z}^{2}]:\frac{\overleftarrow{d}}{d\bar{z}} = 2:[z^{3},\bar{z}]: =  12Hz^{2},  \nonumber \\
&& :[z^{3},\bar{z}^{3}]:\frac{\overleftarrow{d}}{d\bar{z}} = 3:[z^{3},\bar{z}^{2}]: = 36Hz^{2}\bar{z}-88H^{2}z,  
\end{eqnarray}
Hence, we find
\begin{eqnarray}
\frac{\overrightarrow{d}}{dz}:[z^{l},\bar{z}^{m}]: = l:[z^{l-1},\bar{z}^{m}]:,  \quad
:[z^{l},\bar{z}^{m}]:\frac{\overleftarrow{d}}{d\bar{z}} = m:[z^{l},\bar{z}^{m-1}]:. 
\end{eqnarray}
For example, the K\"{a}hler potential ${\cal K}$ for the complex manifold (11) should be normal-ordered,
namely,
\begin{eqnarray}
{\cal K}(z,\bar{z}) &\equiv& \frac{1}{2}:z\bar{z}:, \quad 
g_{z\bar{z}} = \overrightarrow{\partial}_{z}{\cal K}(z,\bar{z})\overleftarrow{\partial}_{\bar{z}}, \quad
\Omega_{\cal K} \equiv ig_{z\bar{z}}dz\wedge d\bar{z}.
\end{eqnarray}
Here, $\Omega_{\cal K}$ is the K\"{a}hler form.
Therefore, we have obtained the simplest example of noncommutative K\"{a}hler geometry.
It is possible for us to insert coordinate-independent Weyl rescaling factor to ${\cal K}$:
\begin{eqnarray}
{\cal K}(z,\bar{z}) &=& \frac{1}{2}e^{2\eta}:z\bar{z}:, \quad \eta;\, {\rm const}.
\end{eqnarray}
However, if $\eta=\eta(z,\bar{z})$, namely if it depends on $z$ and/or $\bar{z}$,
the situation becomes unclear and complicated one for a construction
of a K\"{a}hler manifolds through the usual manner due to $[z,\bar{z}]\ne 0$. 
$g_{z\bar{z}}$ is invariant under the K\"{a}hler transformation 
${\cal K}\to{\cal K}+\psi(z)+\bar{\phi}(\bar{z})$.


Now, we wish to say that, if we introduce the infinite-dimensional closed sets
${\bf T}\equiv\{z^{r}|z\in{\bf C},-1\le r\le 1\}$ and
$\bar{\bf T}\equiv\{\bar{z}^{r}|\bar{z}\in{\bf C},-1\le r\le 1\}$, 
${\bf T}\oplus\bar{\bf T}$ contains the fundamental elements for constructing 
the infinite-dimensional algebra on noncommutative Riemann surfaces 
( except $z^{\pm e}$, $z^{\pm \pi}$, $z^{\pm\gamma}$, $z^{\pm 1/e}$, $z^{\pm 1/\pi}$ and $z^{\pm 1/\gamma}$ 
( $\gamma=0.57721\cdots$, the Euler constant, is included in the range of $r$ of ${\bf T}$, 
though it is impossible to make the exponents of $z^{\pm 1/\gamma}$ by composites of 
any prime or rational numbers )).
We define the subsets of ${\bf T}$ as 
${\bf O}\equiv\{z^{r}|z\in{\bf C},0 < r\le 1\}$, ${\bf S}\equiv\{z^{r}|z\in{\bf C},-1\le r < 0\}$ 
and ${\bf U}\equiv z^{0}=1$. The sets of their complex conjugations are also be given
in the same way of $\bar{\bf T}$.
All of the infinite-dimensional sets ${\bf O}$, ${\bf S}$, ${\bf O}\oplus{\bf U}$, ${\bf S}\oplus{\bf U}$, 
$\bar{\bf O}$, $\bar{\bf S}$, $\bar{\bf O}\oplus{\bf U}$ and $\bar{\bf S}\oplus{\bf U}$
include infinite-number of elements for the subalgebra of algebra given by elements of 
${\bf T}\oplus\bar{\bf T}$.
The elements of all of the cases ${\bf O}$, ${\bf S}$, $\bar{\bf O}$, $\bar{\bf S}$, 
${\bf O}\oplus{\bf U}$, ${\bf S}\oplus{\bf U}$, $\bar{\bf O}\oplus{\bf U}$ and $\bar{\bf S}\oplus{\bf U}$ 
generate commutative subalgebra, while the subalgebra of the ${\bf O}\oplus\bar{\bf O}$, 
${\bf S}\oplus\bar{\bf S}$, ${\bf O}\oplus{\bf S}$, $\bar{\bf O}\oplus\bar{\bf S}$,
${\bf T}$, $\bar{\bf T}$ and ${\bf T}\oplus\bar{\bf T}$
become nontrivial due to the noncommutativity. 
For example, the effect of the noncommutativity arises when we 
take multiplications of $z$ and $z^{-1}$, or the conversion $\varphi: z\leftrightarrow z^{-1}$.
Hence we can mention that, the noncommutativity nontrivially relates a zero with a singular point ( a pole ) 
on the $z$-plane.
It should be noticed that, these algebraic properties do not depend on a specific choice
of a domain $\Omega$ for holomorphic functions, 
they have nothing to do with whether $\Omega$ includes a branch point or not.
The algebraic properties of a flat Riemann surface is the same with that of the noncommutative $z$-plane
where the surfaces are locally determined on a domain of a function $f(z)$.
This fact has the origin in the biholomorphicities of several mappings of Riemann surfaces.
The noncommutative $z$-plane does not satisfy the criteria of the field of complex numbers ${\bf C}$,
because $z$ ( $z^{l}$ ) does not have its inverse in the algebra.
It is also possible for us to generalize the infinite-dimensional set 
for the fundamental elements of the algebra by extending the exponents to a complex domain: 
\begin{eqnarray}
{\cal T} &\equiv& {\cal O}\oplus{\bf U}\oplus{\cal S},    \nonumber \\
{\cal O} &\equiv& \{z^{\zeta}|z,\zeta\in{\bf C},0<|\zeta|\le 1\}, \quad
{\cal S} \equiv \{z^{-\zeta}|z,\zeta\in{\bf C},0<|\zeta|\le 1\},   \nonumber \\ 
{\cal H} &=& [z^{\zeta},\bar{z}^{\bar{\zeta}}] = [z^{\Re\zeta+i\Im\zeta},\bar{z}^{\Re\zeta-i\Im\zeta}], \quad 
0 = [\zeta,\bar{\zeta}].
\end{eqnarray}
It seems interesting for us to investigate algebra under the condition $[\zeta,\bar{\zeta}]\ne 0$.


For the investigation of algebra of ${\bf T}\oplus\bar{\bf T}$, 
we examine several commutators of radical roots.
The commutators of $z^{1/2}$ and $\bar{z}^{1/2}$ will be obtained as follows:
\begin{eqnarray}
&& x^{(2)}_{1} = [z^{1/2},\bar{z}^{1/2}], \quad
(z^{1/2})^{2} = z, \quad (\bar{z}^{1/2})^{2} = \bar{z},  \nonumber \\
&& [z,\bar{z}^{1/2}] = z^{1/2}[z^{1/2},\bar{z}^{1/2}]+[z^{1/2},\bar{z}^{1/2}]z^{1/2} = z^{1/2}x^{(2)}_{1}+x^{(2)}_{1}z^{1/2},  \nonumber   \\
&& [z^{1/2},\bar{z}] = \bar{z}^{1/2}[z^{1/2},\bar{z}^{1/2}]+[z^{1/2},\bar{z}^{1/2}]\bar{z}^{1/2} = \bar{z}^{1/2}x^{(2)}_{1}+x^{(2)}_{1}\bar{z}^{1/2},  \nonumber  \\ 
&& 2H = [z,\bar{z}] = [z^{1/2}z^{1/2},\bar{z}^{1/2}\bar{z}^{1/2}]   \nonumber \\
&& \qquad = z^{1/2}[z^{1/2},\bar{z}^{1/2}]\bar{z}^{1/2} + \bar{z}^{1/2}[z^{1/2},\bar{z}^{1/2}]z^{1/2}
+ [z^{1/2},\bar{z}^{1/2}]z^{1/2}\bar{z}^{1/2} + \bar{z}^{1/2}z^{1/2}[z^{1/2},\bar{z}^{1/2}]  \nonumber \\
&& \qquad = x^{(2)}_{1}z^{1/2}\bar{z}^{1/2} + z^{1/2}x^{(2)}_{1}\bar{z}^{1/2}
+ \bar{z}^{1/2}x^{(2)}_{1}z^{1/2} + \bar{z}^{1/2}z^{1/2}x^{(2)}_{1}.
\end{eqnarray}
Similarly, the algebra of the commutators of $z^{1/4}$ and $\bar{z}^{1/4}$ are 
expanded into the following forms:
\begin{eqnarray}
&& [z^{1/4},\bar{z}^{1/4}] = x^{(2)}_{2}, \nonumber \\
&& [z^{1/4},\bar{z}^{1/2}] = \bar{z}^{1/4}x^{(2)}_{2}+x^{(2)}_{2}\bar{z}^{1/4}, \quad
[z^{1/4},\bar{z}^{3/4}] = \bar{z}^{1/2}x^{(2)}_{2}+\bar{z}^{1/4}x^{(2)}_{2}\bar{z}^{1/4}+x^{(2)}_{2}\bar{z}^{1/2},   \nonumber \\
&& [z^{1/4},\bar{z}] = [z^{1/4},(\bar{z}^{1/4})^{4}] = 4\bar{z}^{3/4}x^{(2)}_{2} \nonumber \\
&& \qquad\qquad = [z^{1/4},(\bar{z}^{1/2})^{2}] 
= \bar{z}^{3/4}x^{(2)}_{2} + \bar{z}^{1/4}x^{(2)}_{2}\bar{z}^{1/2} + \bar{z}^{1/2}x^{(2)}_{2}\bar{z}^{1/4} + x^{(2)}_{2}\bar{z}^{3/4},  \nonumber \\ 
&& \qquad\qquad \to 0 = -3\bar{z}x^{(2)}_{2}+\bar{z}^{1/2}x^{(2)}_{2}\bar{z}^{1/2}
+\bar{z}^{3/4}x^{(2)}_{2}\bar{z}^{1/4} + \bar{z}^{1/4}x^{(2)}_{2}\bar{z}^{3/4},  \nonumber \\
&& [z^{1/2},\bar{z}^{1/4}] = z^{1/4}x^{(2)}_{2}+x^{(2)}_{2}z^{1/4}, \quad 
[z^{3/4},\bar{z}^{1/4}] = z^{1/2}x^{(2)}_{2}+z^{1/4}x^{(2)}_{2}z^{1/4}+x^{(2)}_{2}z^{1/2}, \nonumber \\
&& [z^{1/2},\bar{z}^{3/4}] = [(z^{1/4})^{2},(\bar{z}^{1/4})^{3}] 
= z^{1/4}x^{(2)}_{2}\bar{z}^{1/2} + z^{1/4}\bar{z}^{1/4}x^{(2)}_{2}\bar{z}^{1/4} + x^{(2)}_{2}z^{1/4}\bar{z}^{1/2}   \nonumber \\
&& \qquad\qquad +z^{1/4}\bar{z}^{1/2}x^{(2)}_{2}+\bar{z}^{1/4}x^{(2)}_{2}z^{1/4}\bar{z}^{1/4}+\bar{z}^{1/2}x^{(2)}_{2}z^{1/4}  \nonumber \\
&& \qquad\qquad -(x^{(2)}_{2})^{2}\bar{z}^{1/4}-x^{(2)}_{2}\bar{z}^{1/4}x^{(2)}_{2}
-\bar{z}^{1/4}(x^{(2)}_{2})^{2}, \nonumber    \\
&& \qquad\qquad = [z^{1/2},\bar{z}^{1/2}\bar{z}^{1/4}] 
= x^{(2)}_{1}\bar{z}^{1/4}+\bar{z}^{1/2}(z^{1/4}x^{(2)}_{2}+x^{(2)}_{2}z^{1/4}),   \nonumber \\
&& [z^{3/4},\bar{z}^{1/2}] = [z^{1/2}z^{1/4},\bar{z}^{1/2}] 
= x^{(2)}_{1}z^{1/4}+z^{1/2}(\bar{z}^{1/4}x^{(2)}_{2}+x^{(2)}_{2}\bar{z}^{1/4})    \nonumber \\
&& \qquad\qquad = [(z^{1/4})^{3},(\bar{z}^{1/4})^{2}] 
=z^{1/2}x^{(2)}_{2}\bar{z}^{1/4}+z^{1/4}x^{(2)}_{2}z^{1/4}\bar{z}^{1/4}+z^{1/2}\bar{z}^{1/4}x^{(2)}_{2}   \nonumber \\
&& \qquad\qquad +z^{1/4}\bar{z}^{1/4}x^{(2)}_{2}z^{1/4}
+\bar{z}^{1/4}x^{(2)}_{2}z^{1/2}+x^{(2)}_{2}z^{1/2}\bar{z}^{1/4}  \nonumber  \\
&& \qquad\qquad -(x^{(2)}_{2})^{2}z^{1/4} 
- x^{(2)}_{2}z^{1/4}x^{(2)}_{2}-z^{1/4}(x^{(2)}_{2})^{2},  \nonumber  \\
&& [z^{3/4},\bar{z}^{3/4}] = [z^{1/2}z^{1/4},\bar{z}^{1/2}\bar{z}^{1/4}]    \nonumber \\
&& \qquad\qquad = z^{1/2}\bar{z}^{1/4}x^{(2)}_{2}\bar{z}^{1/4} + z^{1/2}x^{(2)}_{2}\bar{z}^{1/2} 
+ \bar{z}^{1/2}z^{1/4}x^{(2)}_{2}z^{1/4} \nonumber \\
&& \qquad\qquad + \bar{z}^{1/2}x^{(2)}_{2}z^{1/2}
- \bar{z}^{1/2}z^{1/2}x^{(2)}_{2} - x^{(2)}_{1}\bar{z}^{1/4}z^{1/4} + x^{(2)}_{1}x^{(2)}_{2}, \nonumber \\
&& \qquad\qquad = [(z^{1/4})^{3},(\bar{z}^{1/4})^{3}] 
= z^{1/2}x^{(2)}_{2}\bar{z}^{1/2} + z^{1/2}\bar{z}^{1/4}x^{(2)}_{2}\bar{z}^{1/4} 
+ z^{1/2}\bar{z}^{1/2}x^{(2)}_{2} \nonumber \\
&& \qquad\qquad + z^{1/4}x^{(2)}_{2}\bar{z}^{1/2}z^{1/4}
+ x^{(2)}_{2}z^{1/4}\bar{z}^{1/2}z^{1/4} + \bar{z}^{1/4}z^{1/4}x^{(2)}_{2}\bar{z}^{1/4}z^{1/4}  \nonumber \\
&& \qquad\qquad + \bar{z}^{1/4}x^{(2)}_{2}\bar{z}^{1/4}z^{1/2} 
+ \bar{z}^{1/2}x^{(2)}_{2}z^{1/2} + \bar{z}^{1/4}z^{1/4}\bar{z}^{1/4}x^{(2)}_{2}z^{1/4},  \nonumber  \\
&& [z,\bar{z}^{1/4}] = [(z^{1/4})^{4},\bar{z}^{1/4}] =  4x^{(2)}_{2}z^{3/4}   \nonumber \\
&& \qquad\qquad = [(z^{1/2})^{2},\bar{z}^{1/4}] 
= z^{3/4}x^{(2)}_{2}+z^{1/2}x^{(2)}_{2}z^{1/4}+z^{1/4}x^{(2)}_{2}z^{1/2}+x^{(2)}_{2}z^{3/4}, \nonumber \\
&& \qquad\qquad \to 0 = -3x^{(2)}_{2}z+z^{1/2}x^{(2)}_{2}z^{1/2}+z^{3/4}x^{(2)}_{2}z^{1/4}+z^{1/4}x^{(2)}_{2}z^{3/4}, \nonumber \\
&& x^{(2)}_{1} = [z^{1/2},\bar{z}^{1/2}] 
= x^{(2)}_{2}z^{1/4}\bar{z}^{1/4} + z^{1/4}x^{(2)}_{2}\bar{z}^{1/4}
+ \bar{z}^{1/4}x^{(2)}_{2}z^{1/4} + \bar{z}^{1/4}z^{1/4}x^{(2)}_{2},  
\end{eqnarray}
and so forth. In general,
\begin{eqnarray}
&& x^{(2)}_{n} \equiv [z^{1/2^{n}},\bar{z}^{1/2^{n}}],   \nonumber \\
&& x^{(2)}_{n-1} = x^{(2)}_{n}z^{1/2^{n}}\bar{z}^{1/2^{n}}
+ z^{1/2^{n}}x^{(2)}_{n}\bar{z}^{1/2^{n}}+ \bar{z}^{1/2^{n}}x^{(2)}_{n}z^{1/2^{n}}
+ \bar{z}^{1/2^{n}}z^{1/2^{n}}x^{(2)}_{n}. 
\end{eqnarray}
Hence, the structure of the "determination equations" of $x^{(2)}_{n}$ are independent on $n$,
and obey a kind of recursion relations (83), though practically the sequence of these equations cannot
determine the numerical values of $x^{(2)}_{n}$ because they include $z^{1/2^{n}}\bar{z}^{1/2^{n}}$
and in fact they act as operators ( not simple $c$-numbers ).
We can say that, $z$ and $\bar{z}$ just give a starting point of the sequence,
and from this point of view, the set $\{ z^{1/2^{n}},\bar{z}^{1/2^{n}} |n=0,1,2,3,\cdots \}$ 
are ${\it democratic}$ in the determination of the noncommutative algebraic structure. 
If we start the discussions of this work from introducing the noncommutativity
at a specific choice of $z^{1/2^{n}}$ and $\bar{z}^{1/2^{n}}$,
similar algebraic structures and functional properties will be obtained:
If we choose $[z^{1/2^{n}},\bar{z}^{1/2^{n}}]=2H$, and next we take the mapping $z^{1/2^{n}}\to z$
and $\bar{z}^{1/2^{n}}\to\bar{z}$,
the whole of algebra recovers essentially the same one with that of the system given by $[z,\bar{z}]=2H$.
The consideration on the commutators $[z^{1/2^{n}},\bar{z}^{1/2^{n}}]$ ( $n=1,2,3,\cdots$ )
might not have strong meanings for the determination of the algebraic structure, 
though $z^{1/2^{n}}\star\bar{z}^{1/2^{n}}$ are always calculable ( be expanded ) 
through the definition of the star products.
These facts are quite interesting, might relate to the existence of an equivalence class
discussed above, and they might indicate a deep meaning of various quantization schemes,
not only geometrical quantization but also the Heisenberg quantization scheme.  
Similarly, the commutators of $z^{1/3}$ and $\bar{z}^{1/3}$ are found to be
\begin{eqnarray}
&& [z^{1/3},\bar{z}^{1/3}] = x^{(3)}_{1},  \nonumber  \\
&& [z^{2/3},\bar{z}^{1/3}] = z^{1/3}x^{(3)}_{1}+x^{(3)}_{1}z^{1/3}, \quad
[z,\bar{z}^{1/3}] = z^{2/3}x^{(3)}_{1}+z^{1/3}x^{(3)}_{1}z^{1/3}+x^{(3)}_{1}z^{2/3},  \nonumber \\
&& [z^{1/3},\bar{z}] = \bar{z}^{2/3}x^{(3)}_{1}+\bar{z}^{1/3}x^{(3)}_{1}\bar{z}^{1/3}+x^{(3)}_{1}\bar{z}^{2/3}, \quad 
[z^{1/3},\bar{z}^{2/3}] = \bar{z}^{1/3}x^{(3)}_{1}+x^{(3)}_{1}\bar{z}^{1/3},  \nonumber  \\
&& [z,\bar{z}^{2/3}] = z^{2/3}x^{(3)}_{1}\bar{z}^{1/3}+z^{1/3}x^{(3)}_{1}z^{1/3}\bar{z}^{1/3}
+z^{2/3}\bar{z}^{1/3}x^{(3)}_{1}+z^{1/3}\bar{z}^{1/3}x^{(3)}_{1}z^{1/3}
+\bar{z}^{1/3}x^{(3)}_{1}z^{2/3}+x^{(3)}_{1}z^{2/3}\bar{z}^{1/3}  \nonumber  \\
&& \qquad\qquad -(x^{(3)}_{1})^{2}z^{1/3} - x^{(3)}_{1}z^{1/3}x^{(3)}_{1}-z^{1/3}(x^{(3)}_{1})^{2},  \nonumber  \\
&& [z^{2/3},\bar{z}] = z^{1/3}x^{(3)}_{1}\bar{z}^{2/3}+z^{1/3}\bar{z}^{1/3}x^{(3)}_{1}\bar{z}^{1/3}
+x^{(3)}_{1}z^{1/3}\bar{z}^{2/3}+z^{1/3}\bar{z}^{2/3}x^{(3)}_{1}
+\bar{z}^{1/3}x^{(3)}_{1}z^{1/3}\bar{z}^{1/3}+\bar{z}^{2/3}x^{(3)}_{1}z^{1/3}  \nonumber \\
&& \qquad\qquad -(x^{(3)}_{1})^{2}\bar{z}^{1/3}-x^{(3)}_{1}\bar{z}^{1/3}x^{(3)}_{1}-\bar{z}^{1/3}(x^{(3)}_{1})^{2}, \nonumber    \\
&& [z^{2/3},\bar{z}^{2/3}] 
= z^{1/3}x^{(3)}_{1}\bar{z}^{1/3} + \bar{z}^{1/3}x^{(3)}_{1}z^{1/3} 
+ \bar{z}^{1/3}z^{1/3}x^{(3)}_{1} + x^{(3)}_{1}z^{1/3}\bar{z}^{1/3},   \nonumber \\
&& [z,\bar{z}] = 2H = [(z^{1/3})^{3},(\bar{z}^{1/3})^{3}] \nonumber \\
&& = 
z^{2/3}x^{(3)}_{1}\bar{z}^{2/3} + z^{2/3}\bar{z}^{1/3}x^{(3)}_{1}\bar{z}^{1/3} 
+ z^{2/3}\bar{z}^{2/3}x^{(3)}_{1} + z^{1/3}x^{(3)}_{1}\bar{z}^{2/3}z^{1/3}  \nonumber \\
&& + x^{(3)}_{1}z^{1/3}\bar{z}^{2/3}z^{1/3} + \bar{z}^{1/3}z^{1/3}x^{(3)}_{1}\bar{z}^{1/3}z^{1/3}
+ \bar{z}^{1/3}x^{(3)}_{1}\bar{z}^{1/3}z^{2/3} 
+ \bar{z}^{2/3}x^{(3)}_{1}z^{2/3} + \bar{z}^{1/3}z^{1/3}\bar{z}^{1/3}x^{(3)}_{1}z^{1/3},    
\end{eqnarray}
In the above results, 
the structures similar to $[z,\bar{z}^{2}]$, $[z^{2},\bar{z}]$, $[z,\bar{z}^{3}]$, $[z^{3},\bar{z}]$,..., 
repeatedly appear.


The Riemann sphere $\widehat{\bf C}={\bf C}\bigcup\{\infty\}$ is constructed as the ordinary case,
\begin{eqnarray}
\sigma_{1} &=& \frac{\xi_{1}}{1-\xi_{3}}, \quad \sigma_{2} = \frac{\xi_{2}}{1-\xi_{3}}, \quad
\xi^{2}_{1}+\xi^{2}_{2}+\xi^{2}_{3} = 1,  \nonumber \\
z &=& \sigma_{1}+i\sigma_{2} = \frac{\xi_{1}+i\xi_{2}}{1-\xi_{3}},  \quad
\bar{z} = \sigma_{1}-i\sigma_{2} = \frac{\xi_{1}-i\xi_{2}}{1-\xi_{3}}.
\end{eqnarray}
Here, the spherical coordinates is $(\xi_{1},\xi_{2},\xi_{3})$.
Hence, the noncommutativity relation will be expressed as follows:
\begin{eqnarray}
2H &=& [z,\bar{z}] = -\frac{2i}{1-\xi_{3}}[\xi_{1},\xi_{2}].
\end{eqnarray}
Though, it is difficult to interpret $\xi_{3}\to 1$ in the commutator,
as we cannot take $\lim_{z\to 0}[z,\bar{z}]$ in the naive sense.
It is evident from the method of construction, $\widehat{\bf C}$
is compact also in the noncommutative case.


For the integration, by the examinations of the following Riemann sums,
\begin{eqnarray}
\sum^{n}_{k=1}z_{k}\Delta z_{k} &=& \sum^{n}_{k=1}z_{k}(z_{k}-z_{k-1}) = \sum^{n}_{k=1}(z_{k}-z_{k-1})z_{k}     \nonumber \\
&\to& \int_{D} zdz = \int_{D} dzz,   \\
\sum^{n}_{k=1}\frac{1}{z_{k}}\Delta z_{k} &=& \sum^{n}_{k=1}\frac{1}{z_{k}}(z_{k}-z_{k-1})
= \sum^{n}_{k=1}\frac{\bar{z}_{k}}{|z_{k}|^{2}}(z_{k}-z_{k-1})    \nonumber \\
&=& \sum^{n}_{k=1}\frac{1}{|z_{k}|^{2}}(z_{k}\bar{z}_{k}-2H-z_{k-1}\bar{z}_{k}+2H)
= \sum^{n}_{k=1}\frac{1}{|z_{k}|^{2}}(z_{k}-z_{k-1})\bar{z}_{k}  \nonumber \\
&\to& \int_{D} \frac{1}{z}dz = \int_{D} dz\frac{1}{z}, \quad \Bigl( \frac{1}{z} \equiv \frac{1}{|z|^{2}}\bar{z}  \Bigr)
\end{eqnarray}
we have found that, $z$ and $dz$ ( $z^{-1}$ and $dz$ ) commute inside the integration.
Hence, if a function $f(z)$ is given in the form of power series of $z$,
its integration will be done in the ordinary method:
\begin{eqnarray}
\sum^{n}_{k=1}f(z_{k})(z_{k}-z_{k-1}) &\to& \int_{D}f(z)dz.
\end{eqnarray}
It is understood for us that, these Riemann sums will be extended to integrations of arbitrarily curves
through the method of a real-parametric representation of a function $f(z)=f(z(t))$ $(a\le t\le b)$:
\begin{eqnarray}
\int_{\Gamma}f(z)dz &=& \int^{b}_{a}f(z(t))\frac{dz(t)}{dt}dt.
\end{eqnarray}
Several theorems of integration,
i.e., the Cauchy theorem, 
\begin{eqnarray}
f(\alpha) &=& \oint_{C}\frac{dz}{2\pi i}\frac{1}{z-\alpha}f(z) = 
\oint_{C}\frac{dz}{2\pi i}f(z)\frac{1}{z-\alpha},
\end{eqnarray}
the residue theorem,
the Liouville theorem, the Groursat theorem, the Morera theorem, the Taylor theorem, and the Laurent theorem
are also satisfied in functions on the noncommutative $z$-plane.
For example, integrations of several commutators are given as follows:
\begin{eqnarray}
\oint_{C}[z,\bar{z}]dz &=& 0, \nonumber \\
\oint_{C}[z,z^{-1}]dz &=& 2iH|z|^{-1}\int^{2\pi}_{0}e^{i\theta}d\theta = 0,  \nonumber \\
\oint_{C}[z,\bar{z}^{n}]dz &=& 8\pi i H r^{2} \quad (n=2), \quad = 0 \quad (n\ne 2),  \nonumber \\
\oint_{C}[z^{n},\bar{z}]dz &=& 0,  \nonumber  \\
\oint_{C}[z,z^{-n}]dz &=& 8\pi i H r^{-2} \quad (n=2), \quad = 0 \quad (n\ne 2),  \nonumber  \\
\oint_{C}[z^{n},z^{-1}]dz &=& 0,   \nonumber \\
& & \qquad ( C = \{z|z=re^{i\theta},0\le\theta\le 2\pi\}, \quad \forall n\in {\bf N} ).
\end{eqnarray}
A contour integration of an entire function will be done under the usual manner.
On the other hand, $z^{l}z^{-m}$, $z^{-l}z^{m}$, $z^{l}\star z^{-m}$ and $z^{-l}\star z^{m}$ 
( $\forall l,m \in {\bf N}$ ) do not have primitive functions because $z^{l}z^{-m}\ne z^{l-m}$.
For example, $z^{2}z^{-1}\ne z$, $z(z^{-1})\ne 1$, $z(z^{-2})\ne z^{-1}$, so forth.
From the same reason, $z^{l}\bar{z}^{m}$, $\bar{z}^{l}z^{m}$, $z^{l}\star\bar{z}^{m}$ 
and $\bar{z}^{l}\star z^{m}$ do not have primitive functions:
One can take derivatives $\overrightarrow{\partial_{z}}(:z^{l}\bar{z}^{m}:)$
or $(:z^{l}\bar{z}^{m}:)\overleftarrow{\partial_{\bar{z}}}$,
though integrations of $z^{l}\bar{z}^{m}$ ( and thus, $:z^{l}\bar{z}^{m}:$ also )
cannot be determined.
It is an exotic character of noncommutative complex analysis.
The entire function $e^{kz}$ ( $k\in {\bf C}$ ) has its primitive function,
and a Cauchy integration of an entire function $f(z)$ will be done under the usual manner.  
The Green-Stokes formula between a 1-form $\omega_{1}$ and a 2-form $\omega_{2}$ becomes
\begin{eqnarray} 
\int_{D}\omega_{2} = \int_{\partial D} \omega_{1}, \quad
\omega_{1} = zd\bar{z}, \quad \omega_{2} = dz\wedge d\bar{z}.
\end{eqnarray}

\section{Functional Analysis}

For examining the existences of complete sets of functional spaces on the noncommutative $z$-plane,
we will show that the bases of the Fourier transformation are orthogonal:
\begin{eqnarray}
\int^{\infty}_{-\infty} dz e^{ikz}\star e^{-ik'z} &=& \int^{\infty}_{-\infty} dz\Bigl[ e^{i(k-k')z}+H\{e^{ikz},e^{-ik'z}\}^{P.B.}_{z,\bar{z}}+\cdots  \Bigr]     \nonumber \\
&=& \int^{\infty}_{-\infty} dz e^{i(k-k')z} \nonumber \\
&=& \delta(k-k'),
\end{eqnarray}
because the exponential functions are regular and holomorphic.
Here, the integration has been performed over the real axis $-\infty < \Re(z) < +\infty$. 
$\delta(k-k')$ is a distribution called as a Dirac delta function.
This concrete example of the orthonormal relation will make a path toward considerations of functional analysis
in the noncommutative theory.
In fact, a large part of interesting properties of noncommutative Riemann surfaces
arises from characteristics of star products of functions $f(z,\bar{z})$ and $g(z,\bar{z})$.
For examining the structure of functional spaces on the noncommutative Riemann surface, 
we introduce the formal definitions of a $L_{2}[D]$ norm and 
an inner product given by the star products as follows:
\begin{eqnarray}
\| f\|_{\star} &\equiv& \Bigl[ \int_{D} dz \overline{f(z)}\star f(z) \Bigr]^{1/2}, \quad    
\langle f,g\rangle_{\star} \equiv \int_{D} dz \overline{f(z)}\star g(z),  
\end{eqnarray}
or 
\begin{eqnarray}
\| f\|_{\star} &\equiv& \Bigl[ \int_{D} dzd\bar{z} \overline{f(z,\bar{z})} \star f(z,\bar{z}) \Bigr]^{1/2}, \quad     
\langle f,g\rangle_{\star} \equiv \int_{D} dzd\bar{z} \overline{f(z,\bar{z})} \star g(z,\bar{z}),  
\end{eqnarray}
Here, the region of $z$ and $\bar{z}$ integration is defined by an open set $D$,
and $dzd\bar{z}=-2id\sigma_{1}d\sigma_{2}$.
A two-dimensional theory of functional analysis ( integrations will be performed over two-dimensional domains )
is important when we consider (quasi)conformal mapping toward the examination
of Teichm\"{u}ller spaces of Riemann surfaces~[20].  
We assume the integrations are finite, uniquely determined and convergent.
On the other hand, a norm and an inner product of functions under ordinary point product will become
\begin{eqnarray}
\| f\| &\equiv& \Bigl[ \int_{D} dzd\bar{z} \overline{f(z,\bar{z})}f(z,\bar{z}) \Bigr]^{1/2}, \quad     
\langle f,g\rangle \equiv \int_{D} dzd\bar{z} \overline{f(z,\bar{z})}g(z,\bar{z}). 
\end{eqnarray}
Thus, 
\begin{eqnarray} 
\| f\|^{2}_{\star} &=& \| f\|^{2} + \int_{D}dzd\bar{z} \Bigl[ 
H \{ \bar{f},f\}^{P.B.}_{z,\bar{z}} + \cdots \Bigr],  \nonumber \\
\langle f,g \rangle_{\star} &=& \langle f,g \rangle + \int_{D} dzd\bar{z} \Bigl[ 
H \{ \bar{f},g\}^{P.B.}_{z,\bar{z}} + \cdots \Bigr].
\end{eqnarray}
We assume $\| f\|$ and $\langle f,g\rangle$ are uniquely obtained and converged.
These differences given above might not finite if power series of the star products do not converge. 
The parameter of the noncommutativity $H$ parametrizes the formal differences of the norm and inner products,
and they will be given as infinite-order series of $H$.
We could not neglect the possible differences of the following quantities:
\begin{eqnarray}
\Delta\|f\|_{\star} &\equiv& 
\Bigg[\int_{D}dzd\bar{z}\overline{f(z,\bar{z})} \star f(z,\bar{z})\Bigg]^{1/2} 
- 
\Bigg[\int_{D}dzd\bar{z}f(z,\bar{z}) \star \overline{f(z,\bar{z})}\Bigg]^{1/2},    
\nonumber \\
\Delta\|f\| &\equiv& 
\Bigg[\int_{D}dzd\bar{z}\overline{f(z,\bar{z})} f(z,\bar{z})\Bigg]^{1/2} 
- 
\Bigg[\int_{D}dzd\bar{z}f(z,\bar{z}) \overline{f(z,\bar{z})}\Bigg]^{1/2}, 
\end{eqnarray}
and thus, we should employ a convention for the definition of the norm $\|f\|_{\star}$ and $\|f\|$.


Now, we prepare a vector space ${\bf H}$ ( $\dim{\bf H}=n$ ) of several vectors 
denoted as ${\cal U}$, ${\cal V}$, ${\cal W}\in {\bf H}$, i.e., 
${\cal V}(z,\bar{z})\equiv[V_{1}(z,\bar{z}),V_{2}(z,\bar{z}),\cdots,V_{n}(z,\bar{z})]$, etc. 
( ${\it Lineare}$ ${\it Mannigfaltigkeit}$~[27] ), where $x{\cal V}+y{\cal W}\in {\bf H}$ 
( $x,y\in{\bf C}$ ).
Products of vectors on ${\bf H}$ will be defined through the star products, and satisfy
the following linear algebra:
\begin{eqnarray}
{\cal V}^{\dagger}\star{\cal W} &\equiv& \sum^{n}_{i=1}\bar{V}_{i}\star W_{i} 
= {\rm tr}\Bigl(\bar{\cal V}\star{\cal W}^{T}\Bigr) 
\ne {\cal W}^{T}\star\bar{\cal V},  \nonumber \\
U(x{\cal V}+y{\cal W}) &=& xU{\cal V}+yU{\cal W} = x{\cal V}'+y{\cal W}',  \nonumber \\
& & {\cal V} = \{V_{n}\}, \quad {\cal W} = \{W_{n}\}, \quad
{\cal V}' = \{V'_{n'}\}, \quad {\cal W}' = \{W'_{n'}\},   \nonumber \\
{\cal V'}^{\dagger}\star{\cal W'} &=& 
({\cal V}^{\dagger}U^{-1})\star(U{\cal W}) = {\cal V}^{\dagger}\star{\cal W},  \nonumber  \\
{\cal U}^{\dagger}\star({\cal V}+{\cal W}) &=& 
{\cal U}^{\dagger}\star{\cal V} + {\cal U}^{\dagger}\star{\cal W},  \nonumber \\
\langle x{\cal U}+y{\cal V}, {\cal W}\rangle_{\star} &=& 
\bar{x} \langle {\cal U}, {\cal W}\rangle_{\star}+ \bar{y} \langle {\cal V}, {\cal W}\rangle_{\star},   \nonumber \\
(x{\cal U})^{\dagger}\star(x{\cal U}) &=& |x|^{2}{\cal U}^{\dagger}\star{\cal U}. 
\end{eqnarray}
A unitary transformation matrix $U$ ( the dimension of $U$ depends on $\{V_{n}\}$ ) 
for the star-product given above must not depend on both $z$ and $\bar{z}$,
and the unitary operator $U$ becomes a bijection between two vector spaces if $n=n'$.
Here, $U^{-1}\star U = U^{-1}U = UU^{-1} = U\star U^{-1} = \hat{1}_{n}$, where $\hat{1}_{n}$
is an identity operator: $\hat{1}_{n}{\cal V}={\cal V}$.
The following relations will be found from the definition for products of vectors:
\begin{eqnarray}
\| {\cal V} \|_{\star} &\equiv& \Bigl( \langle {\cal V}, {\cal V}\rangle_{\star} \Bigr)^{1/2}
= \Bigl( \int_{D} dzd\bar{z} {\cal V}^{\dagger}\star{\cal V}\Bigr)^{1/2},  \nonumber \\ 
\Bigl( \| {\cal U}+{\cal V} \|_{\star} \Bigr)^{2} &=& 
\Bigl( \| {\cal U} \|_{\star} \Bigr)^{2} + \Bigl( \| {\cal V} \|_{\star} \Bigr)^{2} +
\langle {\cal V}, {\cal U}\rangle_{\star} + \langle {\cal U}, {\cal V} \rangle_{\star},  \nonumber  \\
\| x{\cal U} \|_{\star} &=& |x| \| {\cal U} \|_{\star}.
\end{eqnarray} 
$\|{\cal V}\|_{\star}$ denotes a "norm" of a vector determined through the star product.
$\|{\cal V}\|_{\star}$ is a complex number in general.
The Pappus identity is satisfied:
\begin{eqnarray}
(\| {\cal U} + {\cal V} \|_{\star})^{2}+(\| {\cal U} - {\cal V} \|_{\star})^{2} 
&=& 2(\| {\cal U} \|_{\star})^{2}+2(\| {\cal V} \|_{\star})^{2}.
\end{eqnarray}
Hence the distance between ${\cal V}$ and ${\cal W}$ is defined by
\begin{eqnarray}
\| {\cal V} - {\cal W} \|_{\star} \quad ( = \| {\cal W} - {\cal V} \|_{\star} ). 
\end{eqnarray}
In general,
\begin{eqnarray}
\Bigl( \| {\cal U} \star{\cal V} \|_{\star} \Bigr)^{2} &\ne& 
\Bigl( \| {\cal U} \|_{\star} \Bigr)^{2} \Bigl( \| {\cal V} \|_{\star} \Bigr)^{2},
\end{eqnarray}
and the Cauchy-Schwarz inequality 
$\|{\cal U}\star{\cal V}\|^{2}_{\star}\le\|{\cal U}\|^{2}_{\star}\|{\cal V}\|^{2}_{\star}$
does not realize in general ( under our formal discussion ). 
${\bf H}$ can be called as a pseudo Hermite space ( the Hermite inner product is defined ),
though $\langle {\cal V},{\cal W}\rangle_{\star} \ne \overline{\langle {\cal W},{\cal V}\rangle_{\star}}$,
namely the Hermiticity is broken in general.
If a skew symmetry $\langle{\cal V},{\cal W}\rangle_{\star}=-\langle{\cal W},{\cal V}\rangle_{\star}$ realizes,
the inner product has a similar structure to a case of symplectic manifold.
The concepts of strong convergence $\lim_{n\to\infty}\|{\cal V}_{n}-{\cal V}\|_{\star}\to 0$ and
weak convergence $\lim_{n\to\infty}\langle{\cal W},{\cal V}_{n}-{\cal V} \rangle_{\star}\to 0$ 
can be introduced in our theory.
If a vector always satisfies $\| {\cal V}\|_{\star} \ge 0$ when $\|{\cal V}\|_{\star}$ is real, 
and a Cauchy sequence converges ( complete ), namely
$\| {\cal V}_{n} -{\cal V}_{m}\|_{\star}\to 0$ ( $n,m\to\infty$ ) for $\{{\cal V}_{n}\}^{\infty}_{n=1}$,
and in addition, if
\begin{eqnarray}
\overline{\langle {\cal V},{\cal W}\rangle_{\star}} &=& \langle {\cal W},{\cal V}\rangle_{\star}
\end{eqnarray}
is satisfied, the normed vector space becomes a Hilbert space 
( a Banach space of norm defined from the inner product ).
If a mapping $F:{\cal V}\to{\bf C}$ with
\begin{eqnarray}
F[x{\cal V}+y{\cal W}] &=& xF[{\cal V}]+yF[{\cal W}],
\end{eqnarray}
$F$ is called as a linear functional.
If both $F_{1}$ and $F_{2}$ are linear functionals,
\begin{eqnarray}
(xF_{1}+yF_{2})[{\cal V}] &=& xF_{1}[{\cal V}] + yF_{2}[{\cal V}]
\end{eqnarray}
is satisfied.
The formal definition of norm of a linear functional $F$ is given by
\begin{eqnarray}
\|F\|_{\star} &\equiv& \sup_{0\ne{\cal V}\in{\bf H}}\frac{\|F[{\cal V}]\|_{\star}}{\|{\cal V}\|_{\star}}.
\end{eqnarray}
This definition is applicable when $\|F[{\cal V}]\|_{\star}$ and $\|{\cal V}\|_{\star}$ are positive definite.
A linear operator ${\cal A}$ and its adjoint are also be defined,
\begin{eqnarray}
\langle {\cal V}, {\cal A}\star{\cal W}\rangle_{\star} &=& 
\int_{D} dzd\bar{z} [ {\cal V}^{\dagger}\star({\cal A}\star{\cal W}) ], 
\end{eqnarray}
and a self-adjoint case is
\begin{eqnarray}
\langle {\cal V},{\cal A}\star{\cal W}\rangle_{\star} &=& \langle {\cal A}\star{\cal V}, {\cal W}\rangle_{\star}.  
\end{eqnarray}
If ${\cal A}$ is self-adjoint, $\langle {\cal V},{\cal A}\star{\cal V}\rangle$ is real.
For example, a simple scalar function $f(z,\bar{z})$ with
${\cal V}^{\dagger}\star(f\star{\cal W})=(f^{\dagger}\star{\cal V}^{\dagger})\star{\cal W}$ 
can satisfy the self-adjoint condition. 
The norm of ${\cal A}$ is defined by
\begin{eqnarray}
\|{\cal A}\|_{\star} &\equiv& 
\sup_{0\ne {\cal V}\in{\bf H}}\frac{\|{\cal A}\star{\cal V}\|_{\star}}{\|{\cal V}\|_{\star}}
\end{eqnarray}
Moreover, if we have a orthonormal complete set of a linear independent unit vector bases 
${\bf e}_{i}(z,\bar{z})\in{\bf H}$, any vectors will be expressed in the following way:
\begin{eqnarray}
{\bf v}(z,\bar{z}) &=& \sum^{N}_{i=1}v_{i}{\bf e}_{i}(z,\bar{z}), \quad N \in {\bf N} \quad 
{\rm or} \quad N = \infty \quad ( {\rm separable}), \\
\langle {\bf e}_{i}(z,\bar{z}),{\bf e}_{j}(z,\bar{z}) \rangle_{\star} &=&
\int_{D} dzd\bar{z}  \overline{{\bf e}_{i}(z,\bar{z})} \star {\bf e}_{j}(z,\bar{z}) = \delta_{ij}, 
\end{eqnarray}
then
\begin{eqnarray}
v_{i} &=& \langle {\bf e}_{i}, {\bf v}\rangle_{\star}.
\end{eqnarray}
In fact, a complete set of the commutative case can be used also in the noncommutative theory
through a bijective mapping $\varphi:M_{c}\to M_{nc}$,
though we might have to employ a "renormalization" for norm of the bases,
and sometimes we may have to employ the Schmidt orthogonalization procedure.
We have shown that, several notions of functional analysis can be introduced into a functional space
defined on a noncommutative Riemann surface.
It is a clear fact that, if all of components of a vector ${\cal V}$ are given by entire functions of $z$,
the character of the norm and inner products can satisfy the criteria of the Banach-Hilbert topological space.


We consider the following Fredholm-type integral equation of second kind~[28]:
\begin{eqnarray}
u(z,\bar{z}) &=& f(z,\bar{z}) + \lambda\int_{D}dz'd\bar{z}' K(z,\bar{z};z',\bar{z}')\star u(z',\bar{z}').  
\end{eqnarray}
This integral equation will be analyzed after transforming into the form of Neumann series:
\begin{eqnarray}
u(z,\bar{z}) &=& f(z,\bar{z}) 
+ \lambda\int_{D}dz'd\bar{z}' K(z,\bar{z};z',\bar{z}')\star f(z',\bar{z}')  \nonumber \\
& & + \lambda^{2}\int_{D}dz'd\bar{z}' \int_{D}dz''d\bar{z}'' 
K(z,\bar{z};z'',\bar{z}'')\star K(z'',\bar{z}'';z',\bar{z}')\star f(z',\bar{z}') + \cdots  \nonumber \\
&=& \Bigg[  
1 + \lambda\int_{D} dz'd\bar{z}' K
+ \lambda^{2}\int_{D} dz'd\bar{z}' \int_{D} dz''d\bar{z}'' 
K \star K + \cdots \Bigg]\star f.
\end{eqnarray}
The integral operator appear in the equation is a linear operator, and we use the symbolized notation:
\begin{eqnarray}
\hat{T}f[z,\bar{z}] &=& \int_{D} dz'd\bar{z}' K(z,\bar{z};z',\bar{z}')\star f(z',\bar{z}').
\end{eqnarray}
Here, we wish to introduce the following definition for $L_{2}[D]$ norm of the linear operator $\hat{T}$:
\begin{eqnarray}
\| \hat{T} \|_{\star} &\le& R^{-1}_{conv} \equiv 
\sup_{z\in D}\Bigl( \int_{D} dz'd\bar{z}' 
\overline{K(z,\bar{z};z',\bar{z}')} \star K(z,\bar{z};z',\bar{z}')  \Bigr)^{1/2}
\end{eqnarray}
As we have discussed on $\Delta\|f\|_{\star}$, the definition of $\|\hat{T}\|_{\star}$
also has an ambiguity, and the definition given above have taken a convention.
Then the integral equation is written down in the operator expression:
\begin{eqnarray}
u &=& (1+\lambda\hat{T}+\lambda^{2}\hat{T}^{2}_{\star}+\cdots)\star f 
= \sum^{\infty}_{k=0}(\lambda^{k}\hat{T}^{k}_{\star})\star f,  \nonumber \\
\hat{T}^{2}_{\star} &=& \hat{T}\star\hat{T}, \quad 
\hat{T}^{k}_{\star} = \underbrace{\hat{T}\star\hat{T}\star\cdots\star\hat{T}}_{k}.
\end{eqnarray}
If $\hat{T}$ is bounded, $\|\hat{T}\|_{\star}$ has an upper limit as 
$\|\hat{T}\star f\|_{\star}\le c\|f\|_{\star}<+\infty$.
For using this inequality, $\|\hat{T}\star f\|_{\star}$ and $\|f\|_{\star}$ should be real numbers ( $\in {\bf R}$ ).
We also assume $\|\hat{T}^{k}_{\star}\|_{\star}\le\|\hat{T}\|^{k}_{\star}$, where,
\begin{eqnarray}
\|\hat{T}^{k}_{\star}\|_{\star} &=& \underbrace{\|\hat{T}\star\hat{T}\star\cdots\star\hat{T}\|_{\star}}_{k}   \nonumber \\
&=& \sup_{z\in D}\Bigg\{ \int_{D}\int_{D}\cdots\int_{D} \prod^{k}_{j=1} dz_{j}d\bar{z}_{j} \nonumber \\
& & \quad \times \Bigl[ \overline{ 
K(z,\bar{z};z_{1},\bar{z}_{1})\star K(z_{1},\bar{z}_{1};z_{2},\bar{z}_{2})\star\cdots\star
K(z_{k-1},\bar{z}_{k-1};z_{k},\bar{z}_{k}) } \Bigr]    \nonumber \\
& & \quad \times \star
\Bigl[ K(z,\bar{z};z_{1},\bar{z}_{1})\star K(z_{1},\bar{z}_{1};z_{2},\bar{z}_{2})\star\cdots\star
K(z_{k-1},\bar{z}_{k-1};z_{k},\bar{z}_{k}) \Bigr]  \Bigg\}^{1/2}.
\end{eqnarray}
Then we take into account the following inequality,
\begin{eqnarray}
\|\lambda^{k}\hat{T}^{k}_{\star} \|_{\star} \le |\lambda|^{k}\|\hat{T}^{k}_{\star}\|_{\star} 
\le |\lambda|^{k}\|\hat{T}\|^{k}_{\star},
\end{eqnarray}
with properly defined $\|\hat{T}^{k}_{\star}\|_{\star}$.
Then, we find the following condition for the series where it becomes absolutely convergent~[29]:
\begin{eqnarray}
|\lambda| < \| \hat{T} \|^{-1}_{\star}.
\end{eqnarray}
Therefore, the convergence radius is estimated as $R_{conv}$.
The application of the analysis given above to the Dirichlet integral with the Poisson kernel
is an interesting problem for the theory of noncommutative Riemann surfaces.
This is also related to the variation of the Plateau problem~[30],
and it considers the action similar to that of the string theory ( the Nambu-Goto action~[31-34] ).

\section{Conformal Mapping and Teichm\"{u}ller Space}

A one-to-one mapping $\varphi:\{z\to w|z=\sigma_{1}+i\sigma_{2} \in D,w=u+iv \in W \}$ 
becomes conformal when $\varphi'(c)\ne 0$ $(c\in D)$, and satisfies the holomorphicity:
\begin{eqnarray}
\frac{\partial(u,v)}{\partial(\sigma_{1},\sigma_{2})} &=& 
\Bigl(\frac{\partial u}{\partial\sigma_{1}} \Bigr)^{2} + \Bigl(\frac{\partial v}{\partial\sigma_{1}} \Bigr)^{2} 
= |\varphi'(z)|^{2} > 0,  
\quad \Re \varphi(z) = u(z), \quad  \Im \varphi(z) = v(z).
\end{eqnarray}
In the ordinary case, the conformal condition of a mapping $w(t)=\varphi(z(t))$ ( $a\le t\le b$ ) 
is shown by angle of two curves $C:z=z(t)$ and $C':w=w(t)$ given by the arguments of derivatives of 
them as ${\rm arg}w'(t_{0})-{\rm arg}z'(t_{0})={\rm arg}\varphi'(z_{0})$ ( $a\le t_{0}\le b$ ), 
where $w'(t)=dw(t)dt$, $z'(t)=dz(t)/dt$ and $\varphi'(z)=d\varphi(z)/dz$.
By the definition, $\varphi$ has its inverse mapping $\varphi^{-1}$.
If $\varphi$ is holomorphic over a connected open set $D$, and if $\varphi$ is not a constant,
$\varphi(D)$ gives an open set.
When both $\varphi:R\to S$ and $\varphi^{-1}:S\to R$ are holomorphic, 
$R$ and $S$ are biholomorphically ( conformally ) equivalent Riemann surfaces.
These situations realize on the noncommutative $z$-plane with an appropriate $\varphi$.
If $\varphi$ is conformal, $\varphi^{-1}$ is also conformal,
and if $\varphi_{1}$ and $\varphi_{2}$ are conformal, $\varphi_{1}\circ\varphi_{2}$ is also conformal.
The Schwarz's lemma is also be satisfied in the noncommutative case, 
because it is the theorem for entire functions defined on the unit disc ${\cal D}$~[16-18].
( We have seen in the previous section that the algebra and norm of entire functions
will be handled in the same manner with the ordinary case. ) 
The M\"{o}bius transformation for $\widehat{\bf C}$ of the Lie group $SL(2,{\bf C})$ is formally defined 
as $S: z\to w = \frac{az+b}{cz+d}$ ( $a,b,c,d\in {\bf C}$, and the unimodular condition $ad-bc=1$ ).
However, it is not well defined because of the ambiguity coming from 
the noncommutativity of $z$ and $z^{-1}$.
We show two possible choices:
\begin{eqnarray}
w &=& S_{1}(z) \equiv \frac{1}{cz+d}(az+b),  \quad
w = S_{2}(z) \equiv (az+b)\frac{1}{cz+d}, \nonumber \\ 
\frac{1}{cz+d} &=& (cz+d)^{-1},  \quad   S_{1}(z) - S_{2}(z) = [(cz+d)^{-1},az+b] \ne 0.
\end{eqnarray}
The determination equation of fixed points $z=S_{1}(z)$ and $z=S_{2}(z)$ becomes
\begin{eqnarray}
S_{1} : z = \frac{1}{cz+d}(az+b), \quad S_{2} :  z = (az+b)\frac{1}{cz+d}.
\end{eqnarray}
The fixed points of $S_{1}$ and $S_{2}$ trivially coincide as $z=b/(cz+d)$ for the case $a=0$, 
and $z=b/(d-a)$ for the case $c=0$.
The difference of $S_{1}(z)$ and $S_{2}(z)$ is evaluated as follows:
\begin{eqnarray}
(cz+d)^{-1} &=& 
\frac{1}{c}\Bigl\{ \frac{1}{z}+\Bigl(-\frac{d}{c}\Bigr)\frac{1}{z^{2}}+\Bigl(-\frac{d}{c}\Bigr)^{2}\frac{1}{z^{3}} + \cdots  \Bigr\},  \nonumber \\
S_{1}(z)-S_{2}(z) &=& [(cz+d)^{-1},az+b] 
= -\frac{2Ha}{c}\sum^{\infty}_{n=1}n\Bigl(-\frac{d}{c}\Bigr)^{n-1}|z|^{-2n}\bar{z}^{n-1},  \nonumber \\
& & \qquad\qquad\qquad ( |d/c| < |z|, a\ne 0, c\ne 0, d\ne 0 ).   
\end{eqnarray}
Because $S_{1}(z)=S_{2}(z)$ in the commutative case $[z,\bar{z}]=0$, we could say that
this equation indicates the existence of a ${\it degeneracy}$ compared with the noncommutative case
$S_{1}(z)\ne S_{2}(z)$. 
In other words, the noncommutativity $[z,\bar{z}]\ne 0$ lifts the ${\it degeneracy}$
is was not recognized in the ordinary complex analysis. 
Moreover, if we give a linear fractional transformation $w =\frac{az+b}{cz+d}$by an composition of
several mappings,
\begin{eqnarray} 
z &\stackrel{\phi_{1}}{\to}& z + \frac{d}{c},   \quad
z \stackrel{\phi_{2}}{\to} \frac{1}{z},   \quad
z \stackrel{\phi_{3}}{\to} kz,  \quad \Bigl( k \equiv \frac{bc-ad}{c^{2}} \Bigr), \quad
z \stackrel{\phi_{4}}{\to} z + \frac{a}{c} = w,  
\end{eqnarray}
then, $\phi_{4}\circ\phi_{3}\circ\phi_{2}\circ\phi_{1}$ 
gives a successive mapping with no ambiguity, because each "element" of $\phi_{i}$ ( $i=1,2,3,4$ )
has no ambiguity in the definition.
Furthermore, there is the inverse mapping of it as 
$\phi^{-1}_{1}\circ\phi^{-1}_{2}\circ\phi^{-1}_{3}\circ\phi^{-1}_{4}$.
These facts indicate us that, in the noncommutative case we should distinct 
"elementary" and "composite" mappings. 
Another choice for removing the ambiguity is to introduce the definition
\begin{eqnarray}
z \to \frac{a}{c} + \frac{(bc-ad)/c^{2}}{z+d/c}
\end{eqnarray}
for a linear fractional transformation.
( This is the same with $z\to(az+b)/(cz+d)$ in the ordinary case. )
Clearly, this definition is regular at $z\to\infty$.
Because the uniformization theorem is proved by using the linear fractional transformation,
the choice for the definition of it is quite important for us.
We would like to say that, if we use appropriately defined functions for 
the linear fractional transformation and conformal mappings,
the ambiguity coming from the noncommutativity disappears.
The following linear transformation group is an example of 
conformal mapping in the ${\it ordinary}$ case~[16-18,20]:
\begin{eqnarray}
\varphi_{1}: & & z \to z+a, \quad \bar{z} \to \bar{z}+\bar{a}, \quad a\ne 0, 
\quad ( {\rm translation} \, ( {\rm parabolic} )),  \nonumber \\
\varphi_{2}: & & z \to e^{iu}z, \quad \bar{z} \to e^{-iu}\bar{z}, \quad u\in {\bf R}, 
\quad u\ne 2n\pi, \quad n\in {\bf Z}, \quad ( {\rm rotation} \, ( {\rm elliptic} )),  \nonumber \\
\varphi_{3}: & & z \to \lambda z, \quad \bar{z} \to \bar{\lambda}\bar{z}, 
\quad |\lambda|>0, \quad \lambda\ne 1, \quad ( {\rm dilatation} \, ( {\rm hyperbolic} ) ) \nonumber \\
\varphi_{4}: & & z \to \frac{z}{1+\bar{c}z}, \quad \bar{z} \to \frac{\bar{z}}{1+c\bar{z}}, \quad ( {\rm inversion} ).
\end{eqnarray}
Because $u$, $\lambda$ and $a$ are simple $c$-numbers
while $z$ and $\bar{z}$ are regarded as nontrivial noncommutative operators,
\begin{eqnarray}
2H = [z,\bar{z}] 
&\stackrel{\varphi_{1}}{\longrightarrow}& [z+a,\bar{z}+\bar{a}] = [z,\bar{z}],  \nonumber \\
&\stackrel{\varphi_{2}}{\longrightarrow}& [e^{iu}z,e^{-iu}\bar{z}] = [z,\bar{z}],  \nonumber \\
&\stackrel{\varphi_{3}}{\longrightarrow}& [\lambda z,\bar{\lambda}\bar{z}] = |\lambda|^{2}[z,\bar{z}].
\end{eqnarray}
The dilatation acts as a scaling of the noncommutativity parameter $H$:
$|\lambda|\to\infty$ corresponds to the commutative limit.
Any singly-connected open domain $D$ is transformed into the inside of 
${\cal D}$ by a translation and a dilatation.
On the other hand, the inversion $\varphi_{4}$ has an ambiguity and we have four choices
for the definitions: $(1+\bar{c}z)^{-1}z$, $z(1+\bar{c}z)^{-1}$, $1/\bar{c}-\bar{c}^{3}/(\bar{c}z+1)$,
and the decomposition $\varphi_{4}=\phi_{4}\circ\phi_{3}\circ\phi_{2}\circ\phi_{1}$ where
$\phi_{1},\phi_{4}:z\to z+1/\bar{c}$, $\phi_{2}:z\to 1/z$, $\phi_{3}:z\to -z/\bar{c}^{2}$.
The interpretation for these commutators by the idea of ${\it analytisches}$ $Gebilde$ is interesting.
The mapping of a tensor function ( not holomorphic in general ) is given by
\begin{eqnarray}
\varphi_{a}: \Phi(z,\bar{z}) &\to& \Phi(\varphi_{a}(z),\bar{\varphi}_{a}(\bar{z})), \quad a=1,2,3,4.
\end{eqnarray}
It should be emphasized that, the notion of the group $SL(2,{\bf C})$ 
( as same as the notion of the field ${\bf C}$ ) must be changed/extended
because the associativity is broken by the noncommutativity.
For example, $z$ and $-1/z$ ( $\in SL(2,{\bf C})$ ) do not have elements of 
their inverse due to $z(z^{-1})\ne(z^{-1})z\ne 1$.
The automorphisms of Riemann surfaces will become
\begin{eqnarray}
&& z \to az+b, \quad (a,b\in{\bf C}), \quad Aut({\bf C}): {\bf C} \to {\bf C},  \nonumber \\
&& z \to \frac{a}{c} + \frac{(bc-ad)/c^{2}}{z+d/c}, \quad (a,b,c,d\in{\bf C}) \quad 
Aut(\widehat{\bf C}): \widehat{\bf C} \to \widehat{\bf C}, \nonumber \\
&& z \to \frac{a}{c} + \frac{(bc-ad)/c^{2}}{z+d/c}, \quad (a,b,c,d\in{\bf R}) \quad 
Aut({\bf H}): {\bf H} \to {\bf H}.
\end{eqnarray}
Here, we have chosen the functions as the forms without ambiguity.
For example, the mapping ${\bf R}\to{\cal D}$ should be given by 
\begin{eqnarray}
z \to 1-\frac{2i}{z+i},
\end{eqnarray}
while the automorphism of the unit disc ${\cal D}$ should be defined as follows:
\begin{eqnarray}
z \to e^{i\theta}\Bigl[ 1 + \frac{\alpha-\bar{\alpha}^{-1}}{z+\bar{\alpha}^{-1}} \Bigr].
\end{eqnarray}
The Schwarz's lemma guarantees that, $w=e^{i\theta}z$ ( $\theta\in{\bf R}$ ) gives
the mapping $\{z\in{\bf C}|1>|z|\}\to\{w\in{\bf C}|1>|w|\}$ with $z=0\to w=0$.
From these preparations, we conclude that, $\widehat{\bf C}$ is compact while
${\bf C}$ and ${\cal D}\sim{\bf H}$ are noncompact and 
biholomorphically inequivalent ( by the Liouville theorem ).


Let us recall the discussion of conformal field theory in the case of 
ordinary two-dimensional system~[32-35].
A {\it primary} tensor field of conformal weight $(h,\bar{h})$ is transformed under 
conformal transformations ( biholomorphic and bi-antiholomorphic mappings ) 
$z\to z'(z)$, $\bar{z}\to \bar{z}'(\bar{z})$ as
\begin{eqnarray}
\Phi(z,\bar{z}) &=& 
\Bigl(\frac{dz'}{dz}\Bigr)^{h}\Bigl(\frac{d\bar{z}'}{d\bar{z}}\Bigr)^{\bar{h}}\Phi(z',\bar{z}') 
\end{eqnarray}
and the integration measure is transformed as
\begin{eqnarray}
dz'd\bar{z}' &=& \frac{\partial(z',\bar{z}')}{\partial(z,\bar{z})} dz d\bar{z} 
= \Bigl(\frac{dz'}{dz}\Bigr)\Bigl(\frac{d\bar{z}'}{d\bar{z}}\Bigr)dz d\bar{z},
\end{eqnarray}
and thus, the "conformal weight" of the measure is $(-1,-1)$.
In the ordinary case of conformal field theory, these weights are additive under a product of tensors. 
Hence if we introduce the following action functional
\begin{eqnarray} 
\Gamma[\Phi] &=& \int_{D} dz d\bar{z} \Phi(z,\bar{z})
\end{eqnarray}
with the function of weight $(h,\bar{h})=(1,1)$, $\Gamma[\Phi]$ will be called as conformally invariant,
and the noncommutativity might break a part of the invariance of an action under operations of the conformal group.
A famous example of the invariant action is the string model in the conformal gauge~[32-35]:
\begin{eqnarray} 
\Gamma[X]_{0} &=& \int_{\Sigma_{g}} dz d\bar{z} \Bigl[ \partial_{z}X(z,\bar{z})\partial_{\bar{z}}X(z,\bar{z}) \Bigr],
\end{eqnarray}
where, $X(z,\bar{z})$ denotes a string function with the conformal weight $(h,\bar{h})=(0,0)$,
and $\Sigma_{g}$ is a Riemann surface with genus $g$ which determine the domain for the integrand.
For the noncommutative case, the string action might be modified as follows:
\begin{eqnarray} 
\Gamma_{NC}[X] &=& \int_{\Sigma'_{g}} dz d\bar{z} \Bigl[ \partial_{z}X(z,\bar{z})\star \partial_{\bar{z}}X(z,\bar{z}) \Bigr] 
= \Gamma[X]_{0} + H\Gamma[X]_{1} + \frac{H^{2}}{2!}\Gamma[X]_{2} +\cdots,   \\
\Gamma[X]_{1} &=& \int_{\Sigma'_{g}}dzd\bar{z}\{\partial_{z}X,\partial_{\bar{z}}X \}^{P.B.}_{z,\bar{z}} = \int_{\Sigma'_{g}}dzd\bar{z}\Bigl[ 
\partial^{2}_{z}X\partial^{2}_{\bar{z}}X-\bigl(\partial_{z}\partial_{\bar{z}}X\bigr)^{2} \Bigr],   \\
\Gamma[X]_{2} &=& \int_{\Sigma'_{g}}dzd\bar{z}\Bigl[  
\partial^{3}_{z}X\partial^{3}_{\bar{z}}X 
-2\bigl(\partial^{2}_{z}\partial_{\bar{z}}X\bigr)\bigl(\partial_{z}\partial^{2}_{\bar{z}}X\bigr)
+\bigl(\partial_{z}\partial^{2}_{\bar{z}}X\bigr)\bigl(\partial^{2}_{z}\partial_{\bar{z}}X\bigr)
\Bigr],  
\end{eqnarray}
and so forth. Finding the stability condition for the variation with respect to $H$ is an
interesting and important problem for us to construct a string theory on noncommutative Riemann surfaces.
The condition is given by
\begin{eqnarray}
0 &=& \frac{\delta\Gamma_{NC}[X]}{\delta H}\Big|_{H=0} = \Gamma[X]_{1},  \quad
0 < \frac{\delta^{2}\Gamma_{NC}[X]}{\delta H^{2}}\Big|_{H=0} = \Gamma[X]_{2}. 
\end{eqnarray}
Hence, the stationary point is determined by $\Gamma[X]_{1}$,
while the stability is examine by $\Gamma[X]_{n}$ ( ${\bf N}\ni n\ge 2$ ).


The Virasoro algebra will change because it treats functions with including 
both zeros and singularities in general. 
To obtain the Virasoro algebra of the noncommutative $z$-plane, 
we examine conformal generators in the Lie-derivative form.
An infinitesimal conformal transformation to a holomorphic function $\phi$ of 
conformal weight $h$ is defined as follows~[32-35]:
\begin{eqnarray}
\delta_{\varepsilon}\phi(z) &=& 
\Bigl[ \varepsilon(z)\partial_{z} + h(\partial_{z} \varepsilon(z))\Bigr]\phi(z).
\end{eqnarray}
The infinitesimal parameter will be expanded into the modes:
\begin{eqnarray}
\varepsilon(z) &=& \sum_{n\in{\bf Z}}c_{n}z^{n+1},   
\quad  L_{n} \equiv -z^{n+1}\partial_{z},
\quad  \bar{L}_{n} \equiv -\bar{z}^{n+1}\partial_{\bar{z}}, \quad \forall n\in{\bf Z}.
\end{eqnarray}
Now, the Virasoro algebra of the classical level is obtained in the following forms:
\begin{eqnarray}
&& [L_{n},L_{n}] = 0, \quad \forall n \in {\bf Z},  \nonumber  \\
&& [L_{0},L_{2}] = -2L_{2},  \quad  [L_{0},L_{1}] = -L_{1}, \quad  [L_{0},L_{-1}] = L_{-1},    \nonumber \\
&& [L_{0},L_{-2}] = (zz^{-1}+1)L_{-2}+2H|z|^{-2}(L_{-1})^{2},   \nonumber  \\
&& [L_{1},L_{2}] = -L_{3},  \quad  [L_{1},L_{-1}] = 2L_{0},    \nonumber \\
&& [L_{1},L_{-2}] = (z^{2}z^{-2}+2zz^{-1}-4H|z|^{-2})L_{-1}+4H|z|^{-2}L_{0}L_{-1},  \nonumber  \\
&& [L_{2},L_{-2}] = (z^{3}z^{-2}+3z^{-1}z^{2})L_{-1}+6H|z|^{-2}L_{1}L_{-1},   \nonumber \\
&& [L_{2},L_{-1}] = 3L_{1},  \quad [L_{-1},L_{-2}] = L_{-3},  
\end{eqnarray}
and so forth. In the ordinary commutative case, 
$\{L_{-1},L_{0},L_{1}\}$ makes a subalgebra,
and they are the generators of the Lie algebra $sl(2,{\bf C})$ of 
the M\"{o}bius transformation~[35]. In the results obtained above, 
the algebra of $\{L_{-1},L_{0},L_{1}\}$ do not change in spite of the noncommutativity,
and they satisfy the Jacobi identity
$[L_{1},[L_{0},L_{-1}]]+[L_{0},[L_{-1},L_{1}]]+[L_{-1},[L_{1},L_{0}]]=0$.
The similarity transformations of functions $z^{l}$ ( $\forall l\in {\bf N}$ ) become
\begin{eqnarray}
e^{L_{n}}z^{l}e^{-L_{n}} &=& z^{l} + [L_{n},z^{l}] + \frac{1}{2!}[L_{n},[L_{n},z^{l}]] + \cdots,  \nonumber \\
e^{L_{0}}z^{l}e^{-L_{0}} &=& e^{-l}z^{l}, \quad
e^{L_{1}}z^{l}e^{-L_{1}} = e^{-lz}z^{l}, \quad
e^{L_{-1}}z^{l}e^{-L_{-1}} = \sum^{l}_{j=0}\frac{(-l)^{j}}{j!}z^{l-j}.
\end{eqnarray}

Finally, we consider the construction of Teichm\"{u}ller space of noncommtative Riemann surfaces.
We prepare three Riemann surfaces $\tilde{R}$, $R_{c}$ and $R_{nc}$.
$R_{c}$ denotes a Riemann surface of usual commutative number field, while $R_{nc}$ is obtained
by introducing the noncommutativity by a mapping ( morphism ) $\pi_{nc}$.
$\tilde{R}$ is a covering surface of $R_{c}$, and thus $\pi:\tilde{R}\to R_{c}$ ( $\pi$; a projection ).
When $\gamma$ acts to $\tilde{R}$ as $\pi:\tilde{R}\to\tilde{R}$ with satisfying $\pi\circ\gamma=\pi$,
the set of it $\Gamma\equiv\{\gamma\}$ becomes a covering transformation group.
Because of the way of our construction, we have to consider only $\widehat{\bf C}$,
${\bf C}$ and ${\bf H}$ for the covering surface $\tilde{R}$.
Our scheme will be summarized as follows:
\begin{eqnarray}
\tilde{R}/\Gamma \stackrel{\pi}{\longrightarrow} R_{c} \stackrel{\pi_{nc}}{\longrightarrow} R_{nc}.
\end{eqnarray}
We can also consider liftings,
\begin{eqnarray}
\pi(R):\tilde{R} \to R_{c}, \quad \pi_{nc}(R): R_{c} \to R_{nc}, \nonumber \\
\pi(S):\tilde{S} \to S_{c}, \quad \pi_{nc}(S): S_{c} \to S_{nc}, \nonumber \\
\tilde{\phi}:\tilde{R}\to\tilde{S}, \quad \phi:R_{c} \to S_{c}, \quad \phi_{nc}:R_{nc} \to S_{nc}.
\end{eqnarray}
Here, $\tilde{\phi}$ is a lifting of $\phi$, and $\phi$ is a lifting of $\phi_{nc}$.
Hence,
\begin{eqnarray}
\phi\circ\pi(R) = \pi(S)\circ\tilde{\phi}, \quad
\phi_{nc}\circ\pi_{nc}(R) = \phi\circ\pi_{nc}(S), \quad \cdots.
\end{eqnarray}
The Teichm\"{u}ller theory of $\tilde{R}/\Gamma$ and $R_{c}$ is exactly the same with the ordinary case.
Let us explain the ordinary case. 
In the Fuchsian model and the Fuchsian group for the case $\tilde{R}={\bf H}$ with $Aut({\bf H})\in\Gamma$,
the discrete subgroup $\Gamma$ and the fundamental group $\pi_{1}(R)$ of $R_{c}$ is isomorphic~[20]. 
There is a bijective correspondence between the homotopy class of $R_{c}$ and 
the canonical system of generators of the Fuchsian group model $\Gamma$.
Therefore, by using the method of covering surface,
we can construct the Teichm\"{u}ller space $T_{g}$ ( $g\ge 2$ )
and its Fricke coordinates ${\cal F}_{g}:T_{g}\to{\bf R}^{6g-6}$ also in the noncommutative case.
We conclude that, at least by the method of Fuchsian model,
the dimension of the Teichm\"{u}ller space of a noncommutative Riemann surface
is the same with the commutative case.   
Because the mapping ( morphism ) between $R_{c}$ and $R_{nc}$ gives a one-to-one correspondence,
the Riemann mapping theorem and the uniformization theorem also realize in $R_{nc}$
in the sense of correspondences of surfaces,
and the classification of Riemann surfaces by biholomorphicity 
into $\widehat{\bf C}$, ${\bf C}$ and ${\bf H}$ will give the same situation
as the ordinary case.  
For example, any $g=0$ closed ( bounded ) Riemann surfaces is biholomorphically
equivalent to $\widehat{\bf C}$, also in the noncommutative theory.
It should be noticed that, the upper half ${\bf H}$ and the lower half $\overline{\bf H}$
might not equivalent in algebraic relations due to the noncommutativity $[z,\bar{z}]=2H$.
Our discussion of Teichm\"{u}ller theory is just for finding the biholomorphically equivalent relations
of surfaces and domains.

\section{Summary and Prospects}

In this section, we give the summary of this paper, prospects, 
and consider several possible extensions of the main results of this work.
The noncommutative complex analysis has been understood as it has the structure
which reflects a large part of the ordinary case, while it shows various exotic properties
after introducing the noncommutativity into the field ${\bf C}$.
Both of functional analysis and conformal mapping of the noncommutative theory 
has been examined.
We can mention that, there are two important subjects in complex analysis:
One is the conformal mapping and biholomorphicity, while another is the analytic continuation,
and both of them are summarized by the words/concepts of complex differentiable manifolds.
We have discussed these subjects of the noncommutative theory.
On the construction of Teichm\"{u}ller space, the method of Fuchsian group model
with Fricke coordinates has been examined through the consideration on the
universal covering surfaces. 
Other important ways toward Teichm\"{u}ller space
are Fenchel-Nielsen coordinates of geodesic curves of a hyperbolic space by employing 
the pants decomposition of Riemann surfaces, or the examination of quasiconformal mappings~[20].
Similar to the case of conformal mapping examined in this paper,
a quasiconformal mapping also obtains exotic aspects of the noncommutativity $[z,\bar{z}]\ne 0$.
These issues will be considered by the author in other works
to reveal the relation of mapping, surfaces and Teichm\"{u}ller space more clearly.
It is a well-known fact that, considerations on Teichm\"{u}ller space are 
important for the examination of complex dynamics and its fractal structures~[22].
For example, complex dynamics of the Braschke function $f(z)=z^{2}(z+\lambda)/(1+\bar{\lambda}z)$,
and as its results the distinction of Fatou and Julia sets,
obtain the ambiguity discussed in this paper. 
Because the classification of character of fixed points 
( attracting, superattracting, repelling, neutral )
are performed by the examination of norm of mapping function, 
it might also have characteristic aspects of the noncommutativity.


It can be mentioned that, 
the deformation quantization and noncommutative field theories give us
new perspectives on quantum physics.
The Kontsevich deformation quantization formula~[6,8] will be re-expressed by 
the path integral expansion ( or, a partition function in thermal field theory ) 
with the Poisson $\sigma$-model ( a topological sigma model )
and then, the deformation quantization procedure obtains the deep correspondence with quantum field theory~[9,10]. 
In the Kontsevich theory of deformation quantization,
a star product has an equivalence class determined by
$f\star g=D^{-1}(Df\star Dg)$ 
( $Df=f+\sum^{\infty}_{i=1}D_{i}(f)\hbar^{i}$, $D_{i}$ are differential linear operators )~[6,8].
The discussion on the equivalence class relates to a particular case of the Hochshild cocycle and coboundary,
and it has the basis of the realization of the associativity
of a star product and the fulfillment of the axiom of Poisson manifolds. 
Therefore, in the case of this paper and noncommutative field theory,
there might be no the equivalence class in that sense.  
A Poisson manifold is a subspace of multivector fields with fulfilling a Maurer-Cartan equation
expressed by elements of differential graded Lie algebra ( DGLA ),
while an associative star product gives a multidifferential operator space,
be expressed by a DGLA and it has also a Maurer-Cartan equation~[6,8].
The Kontsevich's work gives an identification of a set of equivalent star product
and an equivalent Poisson manifold.
These results of Kontsevich should be generalized in our case,
because both the axiom of Poisson manifolds 
( the skew symmetry of the Poisson brackets, the Jacobi identity, ... )
and associativity of algebra of the star product
are broken due to the introduction of noncommutativity in the base ring ${\bf C}$.
Hence, for example, it seems meaningless to consider a Maurer-Cartan equation
of Poisson tensor fields in our case. 
The application/extension of the methods of path integral expansion of the star products
in our case is also an important problem. 
If a symmetry, for example a rotational symmetry, is broken by the procedure
of deformation quantization, it could be interpreted as a quantum anomaly.
An examination for spontaneous symmetry breaking in the deformation
quantization should be considered as an important problem for us.
Generalizations for including Grassmann numbers and/or supersymmetry, namely investigations
toward the theory of superalgebra of non(anti)commutative super-Riemann surfaces
and super Virasoro algebra give us an interesting issue~[36-40].


Our work of this paper can be understood as a theory of modification of the base ring
( number field ) with deformation quantization.
Very recently, the author have written a paper of the theory of deformation quantization
of functions of quaternions~[41]. The quaternionic number field explicitly breaks the
axiom of a Poisson manifold,
and in that case the algebra of deformation quantization obtain various exotic characters.
It is interesting for us to investigate algebraic aspects of
deformation quantization and noncommutative field theories over 
various number fields ( p-adic~[42], octonions, hypercomplex numbers, ... ). 
This issue might relate the deformation quantization and noncommutative field theory 
to geometric number theory.


The star product for two complex variables $z_{1}$ and $z_{2}$ can be considered,
and there are several choices for noncommutativities of these variables.
If we choose $[z_{1},z_{2}]=H_{2}$, the star product and algebra are obtained through
the analytic continuation $\{\sigma_{a}\} \to \{z_{a}\}$ ( $a=1,2$ ) of functions of $\sigma_{a}$, 
and thus
\begin{eqnarray}
f(z_{1},z_{2})\star g(z_{1},z_{2}) &=& 
f(z_{1},z_{2})\exp\Bigl[i\frac{H_{(2)}}{2}
(\overleftarrow{\partial}_{z_{1}}\overrightarrow{\partial}_{z_{2}}-
\overleftarrow{\partial}_{z_{2}}\overrightarrow{\partial}_{z_{1}})  
\Bigr]g(z_{1},z_{2}).
\end{eqnarray}
This method is regarded as a dualization of spacetime $(z_{1},z_{2})=(z,\tilde{z})$,
similar to the idea of Connes~[1].
Hence, if a path integral expansion of (150) exists, the dynamical system might describe/reflect
the dualized spacetime structure, might relate to 2D quantum gravity.
Algebraic properties of this two-dimensional noncommutative complex manifold ${\bf C}^{2}$
are determined by choices of commutators, and drastically change from our results of this paper.
Especially, choices of commutators for noncommutativities are crucial.
Moreover, the following "conformal" transformations in the above star product can be considered:
\begin{eqnarray}
&&\varrho_{1}: \quad z \to z + \hbar a, \quad \tilde{z} \to \tilde{z}+\hbar\tilde{a},  \nonumber \\
&&\varrho_{2}: \quad z \to \hbar a z, \quad \tilde{z}\to\hbar\tilde{a}\tilde{z},  \nonumber \\
&&\varrho_{3}: \quad z \to e^{i\hbar a}z, \quad \tilde{z} \to e^{i\hbar\tilde{a}}\tilde{z}. 
\end{eqnarray}
The higher-order terms in $\hbar$ will be introduced 
by expanding small parameters $a$ and $\tilde{a}$ in the star product.
Hence, we can say that a deformation of spacetime structure induces 
the deformation of deformation quantization ( generalized deformation quantization ).
Similar to the case of ${\bf C}^{2}$, the star product for functions of several complex variables
will be defined as follows:
\begin{eqnarray}
f(\{z_{n}\}) &=& f(z_{1},z_{2},\cdots,z_{N}),   \quad 
[z_{j},z_{j'}] = H^{jj'}_{(m)},  \quad j,j'=1, \cdots, N, \quad j\ne j' \nonumber \\  
f(\{z_{n}\})\star g(\{z_{n}\}) &=& 
f(\{z_{n}\})\exp\Bigl[ \frac{i}{2}\sum^{N}_{j=1}\sum^{N}_{j'=1}H^{jj'}_{(m)}
(\overleftarrow{\partial}_{z_{j}}\overrightarrow{\partial}_{z_{j'}}-
\overleftarrow{\partial}_{z_{j'}}\overrightarrow{\partial}_{z_{j}})  
\Bigr]g(\{z_{n}\}).
\end{eqnarray}
A starting point of the investigation for infinite-dimensional noncommutative complex manifolds
will be given by the limiting case $N\to\infty$ of the above definitions.
Because the characters of algebra of noncommutative complex manifolds are determined
by the choice for setting commutators of complex coordinates $\{z_{j}\}$,
there are wide variety of the noncommutative complex manifolds,
and classifications for the manifolds are quite interesting and important problems for us.

\end{document}